\documentclass[twocolumn]{jpsj2} 
\newcommand{\Romannum}[1]{\uppercase\expandafter{\romannumeral#1}}
\usepackage{bm}
\usepackage{mathrsfs}
\usepackage{exscale}
\usepackage{color}
\title{To What Extent Iron-Pnictide New Superconductors Have Been Clarified:\\
A Progress Report }
\author{
Kenji \textsc{Ishida}$^{1,2}$\thanks{E-mail address: kishida@scphys.kyoto-u.ac.jp}, Yusuke \textsc{Nakai}$^{1,2}$ \thanks{E-mail address: nakai@scphys.kyoto-u.ac.jp}, and Hideo \textsc{Hosono}$^{3, 4, 5}$ \thanks{E-mail address: hosono@msl.titech.ac.jp}}
\inst{$^{1}$Department of Physics, Graduate School of Science, Kyoto University, Kyoto 606-8502, Japan \\
$^{2}$TRIP (Transformative Research-project on Iron Pnictides), JST (Japan Science and Technology Agency) JST, Sanban-cho bldg., 5, Sanban-cho, Chiyoda, Tokyo 102-0075\\
$^{3}$ERATO-SORST, JST, Frontier Collaborative Research Center, Tokyo Institute of Technology, Yokohama 226-8503\\
$^{4}$Frontier Research Center, Tokyo Institute of Technology, Yokohama 226-8503\\
$^{5}$Materials and Structures Laboratory, Tokyo Institute of Technology, Yokohama 226-8503\\}

\abst{In this review, the authors present a summary of experimental reports on newly discovered iron-based superconductors as they were known at the end of 2008. At the same time, this paper is intended to be useful for experimenters to know the current status of superconductors. The authors introduce experimental results that reveal basic physical properties in the normal and superconducting states. The similarities and differences between iron-pnictide superconductors and other unconventional superconductors are also discussed.  
}

\kword{iron-pnictide, superconductivity, antiferromagnetism, normal-state and superconducting-state properties}

\begin{document}
\maketitle
%
\section{Introduction}
The discovery of the iron-based layered superconductor LaFeAs(O$_{1-x}$F$_x$) reported by Kamihara {\it et al.} on 23rd February 2008 had a great impact to researchers in condensed-matter physics\cite{KamiharaFeAs}. This is partly because the compound containing one of the most familiar ferromagnetic atom ``$iron$'' shows superconductivity at a relatively high temperature of 26 K in LaFeAs(O$_{0.89}$F$_{0.11}$). In addition, Takahashi {\it et al.} reported that the superconducting transition temperature $T_c$ of this compound increases to 43 K under a high pressure of $\sim$ 4GPa\cite{TakahashiNature08}. Soon after these discoveries, researchers reported that the $T_c$ of $R$FeAs(O$_{1-x}$F$_x$) ($R$ = Ce, Pr, Sm, Nd {\it etc}) jumps up to $\sim$ 50 K, which is the highest except for high-$T_c$ cuprates\cite{XHChenSm,ZhiRenNd,ZhiRenPr,PChengGd}.
After these discoveries, many researchers have been interested in these ``{\it iron-pnictide}'' superconductors, and some have been actually studying them and checking cond-mat every day in order to catch up with the progress in research on iron pnictides. The situation is very similar to that when the cuprate superconductor was discovered.
Owing to such competitive research activities, new iron-pnictide superconductors with different crystal structures such as (Ba,K)Fe$_2$As$_2$\cite{RotterPRB2008}, LiFeAs\cite{LiFeAs} and FeSe$_{1-\delta}$\cite{HsuFeSe} have been discovered within a short period. 

All iron-pnictide superconductors include a two-dimensional (2-D) Fe$Pn$ ({\it Pn}: pnictogen atom) layer with a tetragonal structure at room temperature, which is shown in Fig.~1. Therefore, their physical properties are considered to be highly two-dimensional, similarly to those of cuprate, ruthenate and cobaltate superconductors.
\begin{figure}[tbp]
\begin{center}
\includegraphics[width=4.5cm]{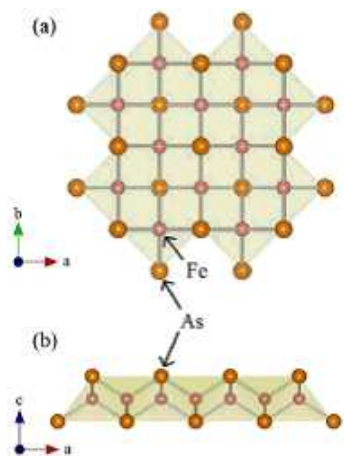}
\end{center}
\caption{(Color online) Top (c-axis) (a) and $b$-axis (b) views of the FeAs layer.  }
\end{figure}

We will try to give a comprehensive review of existing experimental reports on iron-pnictide superconductors. The aim of this study is to review the physical properties of such superconductors in the normal and superconducting states. This article is principally concerned with experiments, and we have not attempted to produce a complete review of theoretical works on iron-pnictide superconductors. 

The content of this article is as follows:
First, we introduce the material variation and phase diagram of iron-pnictide superconductors and review the key experimental results in their normal and superconducting states, particularly for magnetic properties in the normal state and gap structures in the superconducting state. Theoretical models, which are considered to be promising for interpreting iron-pnictide superconductors, are also introduced. On the basis of these experimental results, we summarize their normal and superconducting properties.

\section{Material Variation and Phase diagram}
\begin{figure}[tbp]
\begin{center}
\includegraphics[width=5cm]{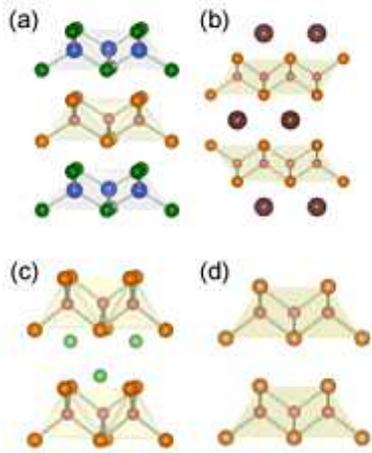}
\end{center}
\caption{(Color online) Crystal structures of (a) LaFeAsO (refered to as ``1111'' in the article), (b) BaFe$_2$As$_2$ (``122''), (c) LiFeAs (``111'') and (d) FeSe (``11''). }
\end{figure}

There are four crystal structures of iron-pnictide superconductors that have been reported so far. These are shown in Fig.~2. The first one is $R$FeAsO with the ZrCuSiAs type structure (space group {\it P4/nmm}), which is shown in Fig.~2(a). This is called the ``1111" structure in this article. It has been reported that approximately 300 $RTPn$O compounds ($R$: rare earth, $T$: late-transition metal, and $Pn$: pnictogen) belong to this type of structure to date\cite{Pottgen}. The $R$O and FeAs layers are stacked along the $c$-axis. The first discovered iron-pnictide superconductor LaFeAs(O$_{1-x}$F$_x$) has this structure, and various rare-earth elements such as Ce, Pr, and Nd can enter the $R$ site. Undoped $R$FeAsO is non-superconducting and shows an antiferromagnetic transition at approximately 150 K. In some compounds, further magnetic anomalies originating from rare-earth moments occur at lower temperatures. Magnetic properties in the undoped ``1111'' structure will be discussed below. Superconductivity emerges when $\sim$ 3\% oxygen is replaced by fluorine ($R$FeAs(O$_{1-x}$F$_x$)) or when oxygen deficiency is introduced ($R$FeAsO$_{1-\delta}$)\cite{KitoJPSJfluorinefree}. Figure 3 shows the F-doping dependence of $T_c$ in LaFeAs(O$_{1-x}$F$_x$)\cite{KamiharaFeAs}, in which anomalies defined in resistivity are also plotted. After superconductivity appears, $T_c$ is nearly unchanged up to $x =$ 0.14. This tendency of $T_c$ has been observed in other $R$FeAs(O$_{1-x}$F$_x$) systems. The maximum $T_c$ of each $R$FeAs(O$_{1-x}$F$_x$), and the F concentration giving the maximum $T_c$ are listed in Table I. The highest $T_c$ = 55 K was reported in SmFeAs(O$_{1-x}$F$_x$)\cite{ZhiRenSm}. 
Quite recently, superconductivity has also been observed in the ``1111'-structure $A$(Fe$_{1-x}$Co$_x$)AsF ($A$ = Ca and Sr) with electron doping. The maximum $T_c$ is 22 K ($A$ = Ca with $x=0.1$)\cite{MatsuishiJACS2008} and 4 K ($A$ = Sr with $x=0.125$)\cite{MatsuishiJPSJ2008}, respectively.      

\begin{figure}[tbp]
\begin{center}
\includegraphics[width=5.5cm]{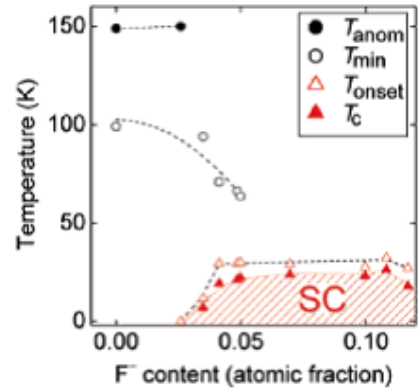}
\end{center}
\caption{(Color online) F-doping dependence of $T_c$ in LaFeAs(O$_{1-x}$F$_x$)\cite{KamiharaFeAs}. Anomalies determined from resistivity are also plotted. Figure reprinted from Y. Kamihara {\it et al.}: J. Am. Chem. Soc. {\bf 130} (2008) 107006. Copyright 2008 by the American Chemical Society. }
\end{figure}
\begin{table}[t]
\caption
{Maximum $T_c$ in each $R$FeAs(O$_{1-x}$F$_x$). The F concentration $x$, which gives the maximum $T_c$ is shown. $T_c^{\rm Max}$ is determined at the onset temperature of superconducting transition in resistivity measurements.}
\begin{tabular}{c|cccccccc} \hline \hline
$R$ & La & Ce & Pr & Nd & Sm & Gd & Tb & Dy\\ \hline 
$T_c^{\rm Max}$ [K] & 28 & 41 & 52 & 52 & 55 & 36 & 46 & 45 \\
$x$ & 0.11 & 0.16 & 0.11 & 0.11 & 0.1 & 0.17 & 0.1 & 0.1 \\ 
Ref.&\cite{KamiharaFeAs} &\cite{GFChenCe} &\cite{ZhiRenPr} &\cite{ZhiRenNd} &\cite{ZhiRenSm} &\cite{PChengGd} &\cite{TbDy} &\cite{TbDy} \\
\hline \hline
\end{tabular}
\label{tableI}
\end{table}

The second iron-pnictide family possesses the ThCr$_2$Si$_2$-type structure, which is shown in Fig.~2(b). This structure is abbreviated as ``122'', and is well-known in ``heavy-fermion (HF)'' compounds, since the first discovered HF superconducting CeCu$_2$Si$_2$\cite{Steglich} and metamagnetic CeRu$_2$Si$_2$ have this structure. The ($R_2$O$_2$)$^{2+}$ layer in $R$FeAsO is replaced by a single divalent ion ($A^{2+}$) layer, and thus the electron count is not altered. $A$Fe$_2$As$_2$ becomes superconducting with hole doping, achieved through the partial substitution of the $A$ site with monovalent $B^+$ (($A_{1-x}B_x$)Fe$_2$As$_2$) ($A$ = Ba, Sr, Ca, $B$ = K, Cs, Na)\cite{RotterPRL2008,SasmalPRL2008}.  Superconductivity also emerges by electron doping via partial substitution of cobalt for iron ($A$(Fe$_{1-x}$Co$_x$)$_2$As$_2$)\cite{SefatPRL2008Co122}. The undoped parent ``122'' compounds show a similar antiferromagnetic ordering as the ``1111'' compounds. Phase diagrams of (Ba$_{1-x}$K$_x$)Fe$_2$As$_2$ and BaFe$_{2-x}$Co$_x$As$_{2}$ were developed from electric resistivity measurements, as shown in Figs.~4(a) and 4(b)\cite{ChenCoexistenceBaKFe2As2,WangPhaseDiagramBaFe2-xCoxAs2}. In these phase diagrams, the coexistence of antiferromagnetic ordering with superconductivity was observed in the ``122'' compounds. However, $\mu$SR measurements clarified the phase separation into the superconducting and magnetic phases in (Ba$_{1-x}$K$_x$)Fe$_2$As$_2$\cite{AczelPRB08} as well as in LaFeAs(O$_{1-x}$F$_x$)\cite{Luetkens,TakeshitaJPSJ08} and Ca(Fe$_{1-x}$Co$_x$)AsF\cite{TakeshitaCaFeAsF}. It was suggested that the volumetric expansion of superconducting domains upon electron doping to the Fe$_2$As$_2$ layers might be understood by the formation of a ``swiss-cheese'' superconducting phase\cite{TakeshitaNJP}.
It was also reported that superconductivity is induced by pressure\cite{AlirezaJPhys2009}, as well as by partial substitution. From the resistivity measurements of SrFe$_2$As$_2$ under pressure, Kotegawa {\it et al.} reported that superconductivity with a sharp resistive transition appears, accompanied by the suppression of the antiferromagnetic state\cite{KotegawaSrFe2As2JPSJ2008}. In contrast, the phase diagram in which $T_N$ covers $T_c$ has been reported from the pressure study of BaFe$_2$As$_2$\cite{FukazawaNMRHP}. The relationship between antiferromagnetism and pressure-induced superconductivity has been debated.

\begin{figure}[tbp]
\begin{center}
\includegraphics[width=6cm]{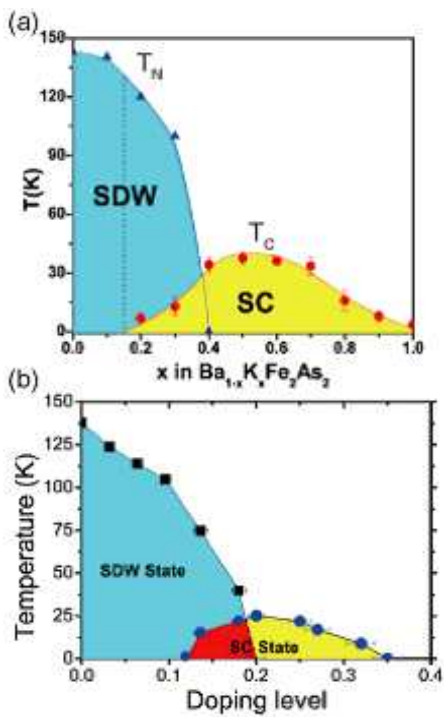}
\end{center}
\caption{(Color online) Phase diagrams of (a) (Ba$_{1-x}$K$_x$)Fe$_2$As$_2$\cite{ChenCoexistenceBaKFe2As2}, (b) BaFe$_{2-x}$Co$_x$As$_2$\cite{WangPhaseDiagramBaFe2-xCoxAs2}. The SDW-type antiferromagnetic transition temperatures $T_N$ and $T_c$ are determined from resistivity measurements. The magnetic state is denoted ``SDW'' in the phase diagram, but is still controversial. The magnetic state of BaFe$_2$As$_2$ is briefly introduced in {\it \S 3.1.3}. Figure reprinted from H. Chen {\it et al.}: EuroPhys. Lett. {\bf 85} (2009) 17006. Copyright 2009 by EDP Sciences.}
\end{figure}

The third structure was reported in superconducting LiFeAs\cite{LiFeAs}, called the ``111'' structure and is shown in Fig.~2 (c). LiFeAs crystallizes into the Cu$_2$Sb-type tetragonal structure containing the FeAs layer with an average iron valence of $+2$ like those for the ``1111'' or ``122'' parent compounds. Wang {\it et al.} reported that superconductivity was observed in Li-deficient compounds (Li$_{1-\delta}$FeAs) synthesized with a high-pressure method. Highest $T_c$ = 18 K was reported in Li$_{0.6}$Fe$_2$As$_2$\cite{LiFeAs}. In contrast, it was reported that single-crystal LiFeAs is not in the magnetically ordered state, but exhibits a superconducting transition at $T_c$ = 18 K, with electron-like carriers\cite{TappPRB2008LiFeAs,PitcherChemCommun2008LiFeAs}. It seems that superconductivity realized in LiFeAs seems to be very sensitive to sample preparation, and that further investigation is needed. It was reported that NaFeAs with the ``111'' structure also shows superconductivity at $T_c \sim 9$ K\cite{NaFeAs}.

Recently, superconductivity has been reported at 8 K in $\alpha$-FeSe compounds with the $\alpha$-PbO-type structure, which is shown in Fig.~2(d). This structure, shortened as the ``11'' structure, has the same planar crystal sublattice equivalent to the FeAs layer in the above three structure\cite{HsuFeSe}. It has been clarified that a clean superconducting phase exists only in those samples prepared with intentional Se deficiency (Se-deficient FeSe is thermodynamically stable in the Fe-Se system). It was reported that the application of pressure raises the onset temperature of superconducting transition as high as 27 K at 1.5 GPa\cite{MizuguchiFeSe}. It was also found that Te substitution for Se enhances $T_c$ up to 15.2 K in FeSe$_{0.5}$Te$_{0.5}$, although FeTe is non-superconducting\cite{YehEPL2008FeSeTl}.  This ``11'' structure has attracted attention owing to its being the simplest among the four structures.

The structural parameters of the four structures are summarized in Table II.
\begin{table}[tbp]
\begin{center}
\caption[Crystallographic data of LaFeAsO, BaFe$_2$As$_2$ (297 K), LiFeAs (room temperature), and
$\alpha$-FeSe$_{0.82}$]{Crystallographic data of
LaFeAsO (measured at 175 K)\cite{KamiharaFeAs,CruzNature2008}, BaFe$_2$As$_2$ (297 K)\cite{RotterPRB2008},
LiFeAs (room temp.)\cite{LiFeAs,TappPRB2008LiFeAs}, and $\alpha$-FeSe$_{0.91}$ (room temp.)\cite{HsuFeSe,MargadonnaFeSe}.}
\label{Crystallographic data}
\begin{tabular}{rrrrr} \hline\hline
& LaFeAsO & BaFe$_2$As$_2$ & LiFeAs & FeSe$_{0.91}$ \\\hline
Space group & $P4/nmm$ & $I4/mmm$ & $P4/nmm$ & $P4/nmm$ \\
$a$ (\AA) & 4.03007 & 3.9625 & 3.7914 & 3.77376 \\
$c$ (\AA) & 8.7368 & 13.0168 & 6.364 & 5.52482 \\
$d_{\rm Fe-Fe}$ (\AA) & 2.84969 & 2.802 & 2.6809 & 2.668 \\
$d_{\rm Fe-As(Se)}$ (\AA) & 2.407 & 2.403 & 2.4204 & 2.38 \\
\hline\hline
\end{tabular}
\end{center}
\label{structural parameters}
\end{table}

\section{Physical Properties}
\subsection{Normal-state results}
In this section, the normal-state properties, particularly of the ``1111'' and ``122'' structures, studied with band calculations and various experimental techniques, are reviewed.

\subsubsection{Electronic structure}
\begin{figure}[tbp]
\begin{center}
\includegraphics[width=6cm]{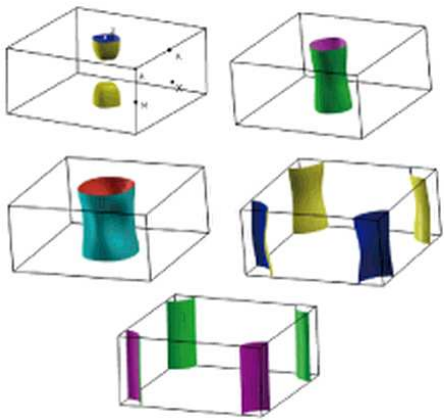}
\end{center}
\caption{(Color online) Fermi surfaces of LaFePO, in which the $\Gamma$ point is centered in the first Brillouin zone\cite{Lebegue}. Figure reprinted from S. Leb\`egue: Phys. Rev. B {\bf 75} (2007) 035110. Copyright 2007 by the American Physical Society.}
\label{Lebegue}
\end{figure}

To understand the bulk electronic properties in the metallic state, it is important to determine electronic structure. After the discovery of superconductivity in LaFePO\cite{KamiharaFeP}, Leb\`egue performed $ab~initio$ calculations of the electronic band structure using the density functional theory\cite{Lebegue}. Leb\`egue showed that the Fermi surfaces of LaFePO consist of five sheets mainly originating from five-fold degenerated Fe $3d$ orbitals: four of them are cylindrical sheets parallel to the $k_z$ direction, in which two hole cylinders are centered along the $\Gamma$ - Z direction while the two electron cylinders are centered along the M - A line, and the fifth sheet consists of a distorted sphere centered at the Z high-symmetry point. The Fermi surfaces reported by Leb\`egue are shown in Fig.~\ref{Lebegue}. Some of the Fermi surfaces were observed experimentally through quantum oscillation experiments\cite{ColdeaLaFePO,SugawaraJPSJ2008}.

\begin{figure}[tbp]
\begin{center}
\includegraphics[width=5.5cm]{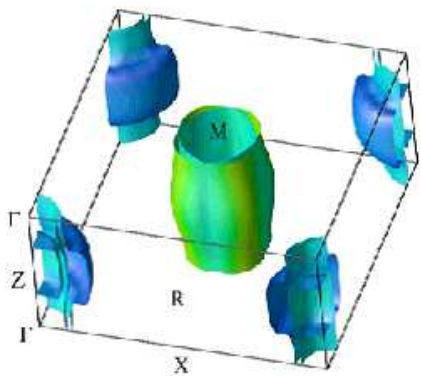}
\end{center}
\caption{(Color online) Fermi surface of LaFeAsO shaded by velocity [darker (blue) is low velocity]\cite{SinghPRL2008}. In this figure, the $\Gamma$ = (0,0,0) is corners in the Brillouin zone, and  Z = (0,0,1/2), X = (1/2,0,0), R = (1/2,0,1/2), M = (1/2,1/2,0) and A = (1/2,1/2,1/2). Figure reprinted from D. J. Singh and M. -H. Du: Phys. Rev. Lett {\bf 100} (2008) 237003. Copyright 2008 by the American Physical Society.}
\end{figure}
The Fermi surfaces of the undoped LaFeAsO were discussed on the basis of the results of the density functional studies by Singh and Du\cite{SinghPRL2008}. They showed that the Fermi surfaces of LaFeAsO is quite similar to those of LaFePO: two high-velocity electron cylinders around the zone edge $M-A$ line, two lower-velocity hole cylinders around the zone center, and an additional heavy 3D hole pocket, which intersects and anticross with the hole cylinders, is centered at $Z$. The heavy 3D pocket is derived from the Fe $d_z$ state, which hybridizes sufficiently with As $p$ and La orbitals to yield a 3D pocket. It is quite interesting that the ground states of LaFePO (superconductivity) and LaFeAsO (stripe antiferromagnet) are quite different, although the Fermi surfaces and band structures near $E_F$ of two compounds are very similar. From the calculation\cite{SinghPRL2008}, it was shown that both compounds are on the border of magnetic instability. Nakamura {\it et al.} discussed the electronic structures of LaFeAsO and LaFePO on the basis of first-principle calculations\cite{NakamuraJPSJ2008}. Although the interactions parameters do not exhibit noticeable differences between the two compounds, they pointed out the difference in bandwidth (the bandwidth of LaFeAsO is slightly ($\sim$20\%) narrower than that of LaFePO) and the difference in a band close to the Fermi surface at the $\Gamma$ point (LaFeAsO has a two-dimensional dispersion with the $d_{x^2-y^2}$ character and LaFePO has a three-dimensional one with the $d_{z^2}$ character).

The electronic structures and Fermi surfaces of the BaFe$_2$As$_2$ ``122'' and LiFeAs ``111'' phases were investigated with the density functional calculations\cite{SinghBaFe2As2PRB2008}. The Fermi surfaces of BaFe$_2$As$_2$ are also similar to those in LaFeAsO, but the hole Fermi surface at the $Z$ point is flattened out, suggesting a more three-dimensional character than that in the ``1111'' and ``111'' structures. 

As an important feature of the Fermi surfaces in these compounds, the cylinders at the $\Gamma$ and M points are nearly nested and can yield strong nesting peaked at ($\pi,\pi$) in the {\it folded} Brillouin zone (two Fe atoms in the unit cell). This can lead in general to enhanced spin fluctuations in this nesting vector and, if these fluctuations are sufficiently strong, they can cause stripe-type SDW ordering. From similar band calculations in other transition-metal pnictides La$T$AsO ($T =$ Cr, Mn, Co and Ni), it was found that the Fermi surface nesting only occurs in LaFeAsO, which makes LaFeAsO distinct from the others\cite{MaPRB08}. Indeed, this stripe-type SDW state appears as the ground state for the parent iron pnictide as discussed in section {\it \S 3.1.3}.    

The main effect of doping is suggested to be a reduction in the degree of nesting of the Fermi surfaces, which seems consistent with the experimental fact that the phase diagram is approximately symmetric against hole doping and electron doping.

\begin{figure}[tbp]
\begin{center}
\includegraphics[width=7cm]{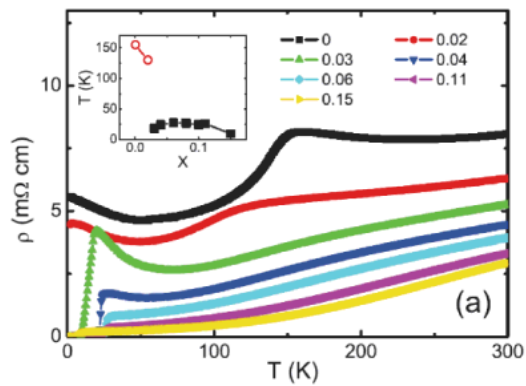}
\end{center}
\caption{(Color online) Temperature dependence of electrical resistivity of LaFeAs(O$_{1-x}$F$_x$). The inset is a phase diagram, in which an anomaly (red circle) in the resistivity and superconducting transition (black square) are shown\cite{DongEPL2008}. Figure reprinted from J. Dong {\it et al.}: EuroPhys. Lett. {\bf 83} (2008) 27006. Copyright 2008 by EDP Sciences.}
\end{figure}
\subsubsection{Transport properties}
Dong {\it et al.} investigated the systematic F-content dependence of the electrical resistivity of LaFeAs(O$_{1-x}$F$_x$), which is shown in Fig.~7\cite{DongEPL2008}. The resistivity of undoped LaFeAsO shows a weak temperature dependence with a high value at high temperatures, and exhibits a steep drop at approximately 150 K with an upturn below 50 K. The resistivity of the 2\% F-doped sample decreases and the 150 K anomaly shifts to a lower temperature and becomes less pronounced. In the 3\% F-doped sample, no anomaly was observed and a superconducting transition occurs at 17 K. With further F-doping, superconducting-transition temperature increases, leading to the highest $T_c \sim 28$ K for 10\% F doping. A quite similar F-doping dependence of $T_c$ was also observed in other $R$FeAs(O$_{1-x}$F$_x$) ($R$ = Ce and Sm) systems\cite{GFChenCe,XHChenSm}. 

\begin{figure}[tbp]
\begin{center}
\includegraphics[width=5.5cm]{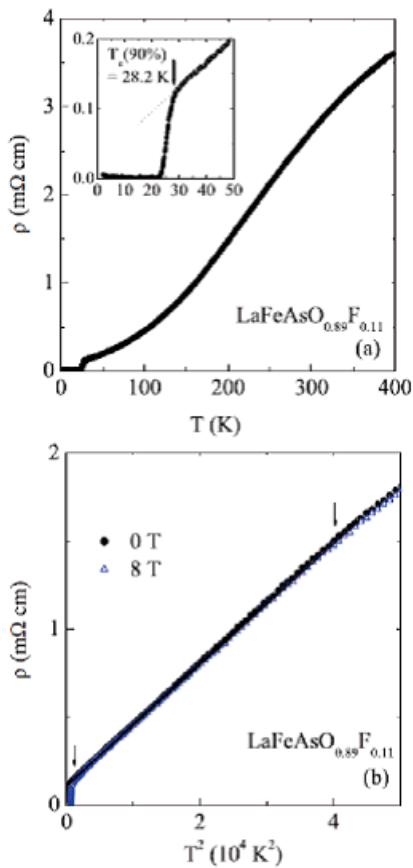}
\end{center}
\caption{(Color online) (a) Temperature dependence of electrical resistivity for LaFeAs(O$_{0.89}$F$_{0.11}$), and (b) $\rho$ plotted against $T^2$ between 0 and $\sim 225$ K. The inset of (a) is the enlarged low-temperature data. In (b), the solid lines represent the linear fit between 35 and 200 K (shown by arrows) for data in 0 and 8 T\cite{SefatPRB2008LaFeAsOF}. Figure reprinted from A. S. Sefat {\it et al.}: Phys. Rev. B {\bf 77} (2008) 174503. Copyright 2008 by the American Physical Society.}
\end{figure}
Using polycrystalline LaFeAs(O$_{0.89}$F$_{0.11}$) with an onset transition temperature $T_c$ = 28.2 K, Sefat {\it et al.} investigated the normal- and superconducting-state properties\cite{SefatPRB2008LaFeAsOF}. The resistivity $\rho$ exhibits a quadratic temperature dependence $\rho = \rho_0+AT^2$ below 200 K with $\rho_0 \sim 0.11$ m$\Omega$~cm, and $A=3.5 \times 10^{-5}$ m$\Omega$cmK$^{-2}$. $A$ is independent of applied magnetic field. The $T^2$ behavior of $\rho$ below 200 K indicates the importance of the umklapp process of electron-electron scattering and is consistent with the formation of a Fermi-liquid state. The $A$ values are comparable to those of semi heavy-fermion compounds such as CePd$_3$ and UIn$_3$\cite{Kadowaki}. 

The temperature dependence of Hall coefficient in LaFeAs(O$_{0.89}$F$_{0.11}$) was measured at a magnetic field of 8 T\cite{SefatPRB2008LaFeAsOF}. The Hall coefficient is negative and weakly temperature-dependent down to 130 K, below which it is approximately constant. If a single band is assumed, the carrier concentrations are $1.7\times 10^{21}$ electrons/cm$^3$ at room temperature, and $1 \times 10^{21}$ electrons/cm$^3$ just above $T_c$, which is consistent with the band calculations\cite{SinghPRL2008}. These calculations also predict the presence of high-velocity electron bands and heavy-hole bands near the Fermi energy, with the electron bands dominating in-plane transport\cite{SefatPRB2008LaFeAsOF}.

\begin{figure}[tbp]
\begin{center}
\includegraphics[width=7cm]{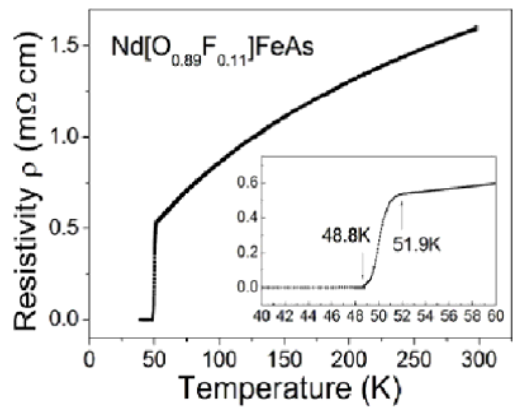}
\end{center}
\caption{Temperature dependence of electrical resistivity of NdFeAs(O$_{0.89}$F$_{0.11}$)\cite{ZhiRenNd}. The inset is the enlarged data around $T_c$. Figure reprinted from Z.-A. Ren {\it et al.}: EuroPhys. Lett. {\bf 82} (2008) 57002. Copyright 2008 by EDP Sciences.  }
\end{figure}
In contrast, no $T^2$ behavior was observed in $R$FeAs(O$_{0.89}$F$_{0.11}$) with $R$ = Nd, Sm, Gd, {\it etc}. Figure 9 shows the temperature dependence of the electrical resistivity of NdFeAs(O$_{0.89}$F$_{0.11}$) with the onset resistivity transition at 51.9 K\cite{ZhiRenNd}. $\rho$ is roughly proportional to temperature down to 200 K, and shows a downward curvature at approximately 150 K. A quite similar temperature dependence was also observed in oxygen-deficient $R$FeAsO$_{1-\delta}$ with $T_c$ greater than 50 K, but the $T^2$ dependence of $\rho$ was reported in PrFeAsO$_{1-\delta}$ with $T_c \sim 35$ K\cite{HashimotoPrFeAsO}. It seems that the $T^2$ behavior is observed in lower-$T_c$ compounds of the ``1111'' system.

\begin{figure}[tbp]
\begin{center}
\includegraphics[width=7cm]{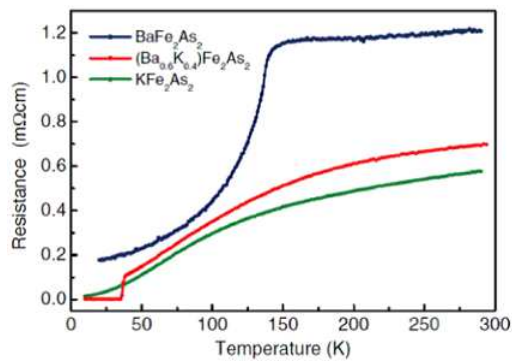}
\end{center}
\caption{(Color online) The temperature dependence of electrical resistivity for BaFe$_2$As$_2$, (Ba$_{0.6}$K$_{0.4}$)Fe$_2$As$_2$ and KFe$_2$As$_2$\cite{RotterPRL2008}. Figure reprinted from M. Rotter {\it et al.}: Phys. Rev. Lett {\bf 101} (2008) 107006. Copyright 2008 by the American Physical Society.}
\label{122rho}
\end{figure}
Figure \ref{122rho} shows the temperature dependence of the electrical resistivity of polycrystalline (Ba$_{1-x}$K$_x$)Fe$_2$As$_2$ ($x$ = 0, 0.4 and 1.0) with the ``122'' structure\cite{RotterPRL2008}. BaFe$_2$As$_2$ has the highest resistance among the three and shows a decrease at 140 K, which originates from SDW and structural anomalies. The resistance of KFe$_2$As$_2$ is considerably smaller and decreases smoothly with temperature, as is typical for a normal metal. The temperature dependence of the resistivity of K-doped (Ba$_{0.6}$K$_{0.4}$)Fe$_2$As$_2$ is similar to that of KFe$_2$As$_2$ and shows no sign of anomaly at approximately 140 K. The resistance drops abruptly to zero at $T_c \sim$ 38 K, which clearly indicates superconductivity. No $T^2$ behavior was observed, but $T$-linear behavior was observed down to $T_c$ below 100 K in (Ba$_{0.6}$K$_{0.4}$)Fe$_2$As$_2$. The resistivity of (Ba$_{0.6}$K$_{0.4}$)Fe$_2$As$_2$ is quite similar to that of $R$FeAs(O$_{1-x}$F$_x$) with higher $T_c$.   

\subsubsection{Magnetic ordering and physical properties of the parent compounds}
As mentioned above, the resistivities of undoped LaFeAsO and BaFe$_2$As$_2$ show an anomaly at approximately 150 K, which is related to the magnetic and structural transitions.

\begin{figure}[tbp]
\begin{center}
\includegraphics[width=7cm]{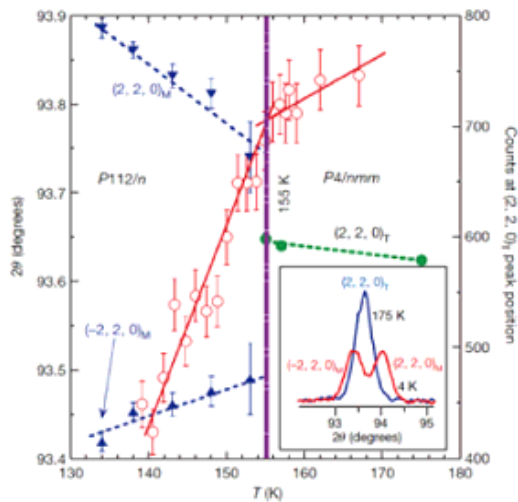}
\end{center}
\caption{(Color online) Temperature dependence of the (2,2,0) nuclear reflection indicative of a structural phase transition at $\sim 155$ K in LaFeAsO. Peak intensities at the (2,2,0)$_T$ (tetragonal) reflection (open symbols, right-hand scale) and position of the (2,2,0)$_T$, (-2,2,0)$_M$ (monoclinic) and (2 ,2,0)$_M$ peaks (solid symbols, left-hand scale) as a function of temperature on cooling. A structural transition from tetragonal symmetry {\it P4/nmm} to monoclinic symmetry {\it P112/n} occurs at $\sim 155$ K. The inset shows the (2,2,0)$_T$ reflection at 175 K and the (-2,2,0)$_M$ and (2,2,0)$_M$ reflections at 4 K\cite{CruzNature2008}. Figure reprinted from C. de la Cruz {\it et al.}: Nature. {\bf 453} (2008) 899. Copyright 2008 by Macmillian Publishers Limited. }
\end{figure}
\begin{figure}[tbp]
\begin{center}
\includegraphics[width=6.5cm]{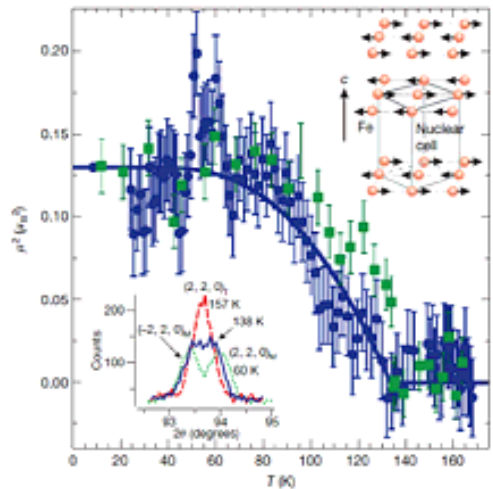}
\end{center}
\caption{(Color online) Temperature dependence of the square of the ordered magnetic moment. The blue circle (green square) denotes experimental data obtained by BT-7 (HB-1A) spectrometer, and the solid line is a simple fit to the mean field theory, which gives $T_N = 137$ K. The bottom-left inset shows the temperature dependence of the nuclear (2,2,0) peak, indicating that the lattice is distorted at 138 K, before the long-range static antiferromagnetic order sets in at $\sim 137$ K. The top-right inset shows the antiferromagnetic structure of the system, giving a $\sqrt{2}a_N\times\sqrt{2}b_N\times~2c_N$ unit cell with moment directions parallel to the planes of iron atoms\cite{CruzNature2008}. Figure reprinted from C. de la Cruz {\it et al.}: Nature. {\bf 453} (2008) 899. Copyright 2008 by Macmillian Publishers Limited. }
\end{figure}

Cruz {\it et al.} reported careful neutron-scattering experiments that reveal the physical properties of the resistive anomaly in LaFeAsO\cite{CruzNature2008}. Figure 11 shows the temperature dependence of the (2,2,0) nuclear reflection, indicative of a structural phase transition at $T_S\sim 155$ K in LaFeAsO. This result demonstrates that LaFeAsO undergoes an abrupt structural distortion below 155 K, and the symmetry changes from tetragonal (space group $P4/nmm$) to orthorhombic (space group $Cmma$) at low temperatures (in the original paper, Cruz {\it et al.} denoted ``monoclinic'' (space group $P112/n$), which becomes the primitive cell of Nomura {\it et al.}'s $Cmma$ space group\cite{Nomura} if the z values of Fe[z(Fe)] and O[z(O)] are assumed to be exactly 0.5 and 0, respectively. See supplemental information\cite{CruzNature2008}). Furthermore, they found the magnetic (1, 0, 3) Bragg peak at 8 K, and an ordered iron moment was estimated to be 0.36(5) $\mu_B$/Fe. They carried out order parameter measurements on the strongest (1,0,3) magnetic peak in order to determine whether or not the ordered magnetic scattering at low temperatures in LaFeAsO is indeed associated with structural-phase transition.  Figure 12 shows the temperature dependence of the square of the ordered moments measured at the Bragg peak. The ordered magnetic moment vanishes at $\sim$ 137 K, approximately $\sim$ 18 K lower than the temperature at which the structural phase transition occurs. The presence of the lattice distortion above the N\`eel temperature $T_N$ is established conclusively in the bottom-left inset of Fig.~12, where a clear lattice distortion is apparent at 138 K. Therefore, the resistive anomaly at approximately 150 K is caused by the structural distortion. In contrast, neither structural nor magnetic anomaly was observed in superconducting LaFeAs(O$_{0.92}$F$_{0.08}$). The disappearance of the static antiferromagnetic order and lattice distortion in the doped superconducting materials suggest that the physical properties of this class of superconductors may have important similarities to those of high-$T_c$ cuprates. 
The top-right inset in Fig.~12 shows a simple stripe-type antiferromagnetic structure within the $ab$-plane doubled along the $c$-axis. It should be noted that the moment of 0.36(5) $\mu_B$ per iron atom is much smaller than the predicted value of $\sim 2.3 \mu_B$ per iron atom, although the magnetic structure is consistent with the theoretical prediction\cite{IshibashiJPSJ2008}. The presence of magnetic frustrations and/or spin fluctuations was suggested to interpret the reduced ordered moment\cite{YildirimPRL2008,QSiPRL2008}. 

Further evidence of the static magnetic ordering at $T_N \sim$ 138 K in LaFeAsO was obtained with $^{57}$Fe M\"ossbauer\cite{KlaussPRL2008,KitaoJPSJ2008}, $\mu$SR\cite{KlaussPRL2008} and $^{139}$La-NMR experiments\cite{NakaiJPSJ2008}. These experiments detected a static internal field and clearly show that magnetic ordering is in a commensurate spin structure. The internal field abruptly develops below $T_N$, which is much steeper than the usual $\sqrt{1-(T/T_N)^2}$ temperature dependence. This steep increase in internal field might originate from the two-dimensionality of magnetic fluctuations, because a similar rapid growth was reported in the cuprate antiferromagnet La$_2$CuO$_4$\cite{MacLaughlinPRL94}.

\begin{figure}[tbp]
\begin{center}
\includegraphics[width=7cm]{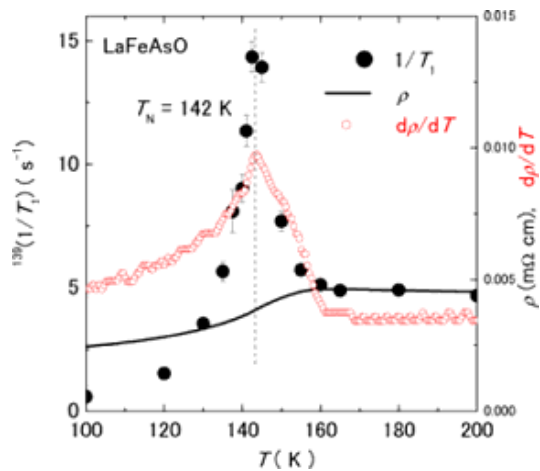}
\end{center}
\caption{(Color online) Temperature dependence of $1/T_1$ of $^{139}$La, resistivity ($\rho$) and $T$ derivative of $\rho$ (d$\rho$/d$T$) in LaFeAsO in the temperature range of $100 - 200$ K.\cite{NakaiJPSJ2008}}
\label{NakaiResT1T}
\end{figure}
From the $^{139}$La-NMR experiment, it is shown that the magnetic ordering is highly related to the structural distortion\cite{NakaiJPSJ2008}. Figure 13 shows the temperature dependence of nuclear spin-lattice relaxation rate $1/T_1$ of La, along with electrical resistivity $\rho$ and its temperature derivative $d\rho/dT$. $1/T_1$ starts to increase below 160 K, where $\rho$ shows a kink owing to the structural transition. The temperature derivative of $\rho$ shows a sharp peak at $T_N$, which coincides with the peak of $1/T_1$. The $1/T_1$ result indicates that magnetic fluctuations start to slow down below 160 K where the structural transition occurs, and magnetic character becomes static below 142 K. The relationship between the structural and magnetic transitions suggests importance of the spin Jahn-Teller effect in LaFeAsO, where the direction of the ordered moments are determined in order to lower magnetic energy after the structural distortion occurs.

Similar structural and magnetic transitions were observed in other rare-earth parent compound $R$FeAsO. The structural transition, antiferromagnetic  transition temperature, and magnetic ordered moments, as well as the ordering temperature of the rare-earth moments and the direction of the rare-earth moments  are summarized in Table III. 

\begin{table*}[t]

\caption
{Structural-transition temperature $T_S$, antiferromagntic ordering temperature $T_N$, and magnetic ordered moment per Fe of undoped compounds with the ``1111'' and ``122'' structures are summarized. In the $R$FeAsO, magnetic ordering temperature by rare-earth magnetic moments ($T_R$) and the direction of the moments ($m_{\rm RE} \parallel$) are also shown. The transition temperatures except that of SrFeAsF were determined by neutron-scattering experiments.}

\begin{tabular}{cc|ccccc} \hline \hline
sample &ref.& $T_S$ [K] & $T_N$ [K] & $m_{\rm ord}$ [$\mu_B$] & $T_R$ [K] & $m_{\rm RE} \parallel$  \\ \hline 
LaFeAsO& \cite{CruzNature2008}& 155 & 137 & 0.36 & &  \\
CeFeAsO&\cite{CeFeAsOPhaseDiagram} & 155 & 140 & 0.83 & 4 & $ab$-plane \\ 
PrFeAsO&\cite{JZhaoPRB2008PrFeAsO} & 153 & 127 & 0.48 & 14 & $c$-axis \\
NdFeAsO&\cite{QiuNdFeAsO} & 150 & 141 & 0.9 & 2 & $ab$-plane \\ 
\hline
CaFeAsF&\cite{MatsuishiJACS2008,XiaoCaFeAsF}& 134 & 114 & 0.49 & & \\
SrFeAsF&\cite{MatsuishiJPSJ2008,bakerSrFeAsF}& 175 & 120 &  & &\\
\hline
CaFe$_2$As$_2$&\cite{GoldmanPRB2008CaFe2As2} & 173 & 173 & 0.8 & & \\
SrFe$_2$As$_2$&\cite{ZhaoPRB2008SrFe2As2,JeschePRB2008SrFe2As2} & 220 & 220 &$0.94-1.0$ & & \\
BaFe$_2$As$_2$&\cite{HuangBaFe2As2} & 140 & 140 & 0.9 & & \\
\hline \hline
\end{tabular}
\label{tableIII}
\end{table*}

\begin{figure}[tbp]
\begin{center}
\includegraphics[width=8cm]{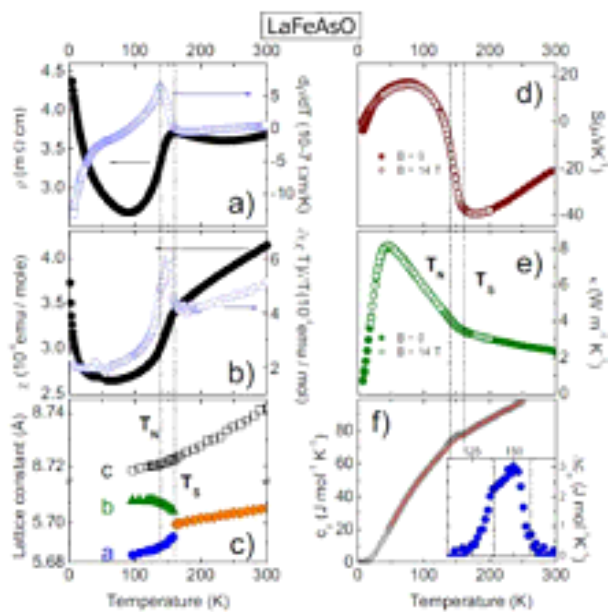}
\end{center}
\caption{(Color online) Temperature dependence of electrical resistivity $\rho$ (a), $\chi = M/B$ (b), lattice constant (c), $\kappa$ (d), $S$ (e) and $c_p$ (f) of the undoped compound LaFeAsO. (a) and (b) also show the derivative $d \rho /dT$ and the magnetic specific heat $d(\chi T)/dT$, respectively. The inset of (f)  displays the anomalous contributions $\Delta c_p$ to the specific heat. The static susceptibility has been measured at an external magnetic field of $B = 1$ T. The dashed lines indicate the temperatures of structural ($T_S \sim 160$ K) and magnetic ($T_N \sim 138$ K) transitions\cite{KondratLaFeAsO}.}
\label{Kondrat}
\end{figure}

Kondrat {\it et al.} reported the temperature dependences of the electrical resistivity $\rho$, the dc magnetic susceptibility, the lattice constants $a$ and $c$, the specific heat $c_p$, the Seebeck coefficient $S$, and the thermal conductivity $\kappa$ of the parent compound LaFeAsO, as summarized in Fig. \ref{Kondrat}\cite{KondratLaFeAsO}. Both phase transitions are visible in $c_p$, which measures the total entropy changes. In the inset of Fig.~\ref{Kondrat}(f), the background-subtracted data $\Delta c_p$ is shown in order to highlight these anomalies. The data confirms a jump in $\Delta c_p$ at $T_S$, which is indicative of a second-order phase transition and an additional anomaly at $T_N$. The similarities between $d(\chi~T)/dT$ and $\Delta c_p$ indicates that the total entropy change connected with both transitions is proportional to that of magnetic origin.  

\begin{figure}[tbp]
\begin{center}
\includegraphics[width=7cm]{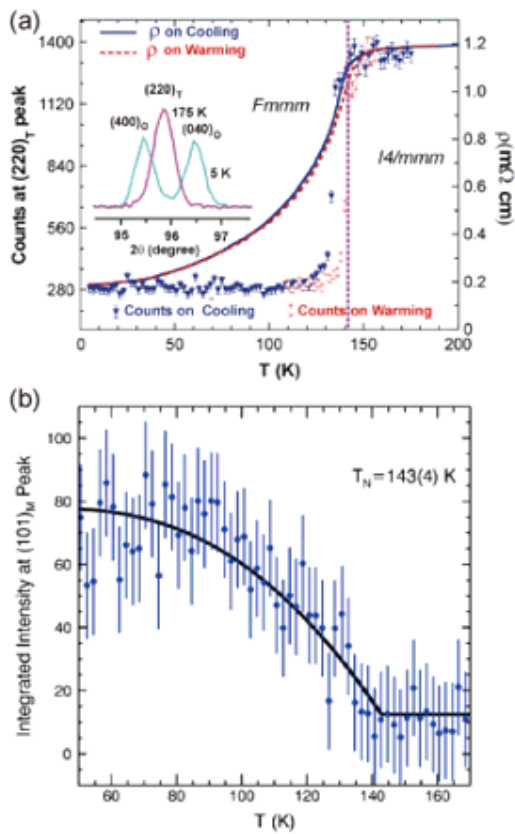}
\end{center}
\caption{(Color online) (a) Temperature dependence of electrical resistivity $\rho$ and the intensity of the neutron diffraction peak at (2,2,0)$_T$ of BaFe$_2$As$_2$. $\rho$ measured on cooling shows a sharp drop at $T_S \sim$ 142 K. The neutron-diffraction intensity decreases abruptly with hysteresis, indicative of the first-order transition. The inset shows the neutron diffraction spectra measured above and below the structural transition. The spectra show the splitting below $T_S$. (b) Temperature dependence of the magnetic Bragg peak (1,0,1)$_M$. The solid line represents the least-squares fit to the mean-field theory\cite{HuangBaFe2As2}. Figure reprinted from Q. Huang {\it et al.}: Phys. Rev. Lett {\bf 101} (2008) 257003. Copyright 2008 by the American Physical Society.}
\end{figure}

Magnetic ordering was also reported in BaFe$_2$As$_2$ with the ``122'' structure from neutron-diffraction measurements\cite{HuangBaFe2As2}. Figure 15(a) shows the temperature dependence of the electrical resistivity and intensity of the neutron diffraction peak at (2,2,0)$_T$. The discontinuity in the intensity of the Bragg reflection and the hysteresis observed during a cooling-and-warming cycle imply that the structural phase transition possesses a first-order character. In addition, a transition to a long-ranged antiferromagnetic state occurs at $T_N \sim 140$ K, where a structural transition from tetragonal to orthorhombic also occurs. The magnetic and structural phase transitions occur simultaneously, which is in contrast to the ``1111''-type compounds. The first-order character of the transition was also suggested from the absence of critical magnetic fluctuations as well as from an abrupt change of the $^{75}$As-NMR intensity with hysteresis obtained in a single-crystal sample\cite{KitagawaBaFe2As2}. The ordered magnetic moment 0.87(3) $\mu_B$ per Fe at 5 K in BaFe$_2$As$_2$ is substantially larger than the saturated moment 0.36(5) $\mu_B$ in LaFeAsO, but is comparable to the moment 0.9 $\mu_B$ in NdFeAsO\cite{ChenPRB2008NdFeAsO} and the moment 1.01 $\mu_B$ in the same ``122'' structure SrFe$_2$As$_2$\cite{JeschePRB2008SrFe2As2}. The first-order structural transition accompanied with the antiferromagnetic transition seems to be a common feature of the ``122'' compounds.  

\begin{figure}[tbp]
\begin{center}
\includegraphics[width=7cm]{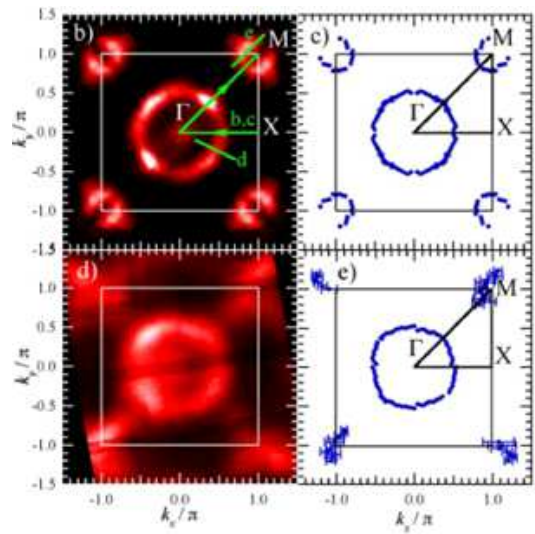}
\end{center}
\caption{(Color online) (b) Measured Fermi surface of NdFeAsO$_{0.9}$F$_{0.1}$ with $T_c \sim 53$ K. Intensity of the photoelectrons integrated over 20 meV at the chemical potential obtained with 22 eV photons at $T = 70$ K. (c) The areas of color mark the locations of the Fermi surfaces. (d,e) The same data as those in (b,c) but obtained with 77 eV photons\cite{LiuNdFeAsOF}. }
\label{LiuNdFeAsOF}
\end{figure}
\subsubsection{Fermi surface experiments} 
Angle resolved photoemission spectroscopy (ARPES) is a powerful technique for investigating electronic properties. Liu {\it et al.} performed ARPES experiments on large-grain superconducting NdFeAsO$_{0.9}$As$_{0.1}$ ($T_c \sim 53$ K), and reported that well-defined Fermi surfaces consist of a cylindrical large hole pocket at the Brillouin zone center $\Gamma$ (0,0,0) and a smaller electron pocket at each corner of the Brillouin zone (M points), as shown in Fig.~\ref{LiuNdFeAsOF}\cite{LiuNdFeAsOF}. Although some band calculations yield an ellipsoidal Fermi surface around Z (0,0,1), this was not observed for the studied photon energies. The overall location and shape of the Fermi surface reasonably agree with band calculations; however, the band dispersion is quite complicated with many flat bands located just below the chemical potential. 

The Fermi surfaces of Ba$_{1-x}$K$_x$Fe$_2$As$_2$ with the ``122'' structure have been investigated by several groups.\cite{HDingBaKFe2As2, ZhaoBaKFe2As2, CLiuPRL2008Ba122, HLiuPRB2008ARPES, YangARPESparentBa122, YZhangARPESSrK122, DingARPESBaK122,SatoARPESK122} It was found that the Fermi surfaces of the undoped BaFe$_2$As$_2$ consist of hole pockets at $\Gamma$ (0,0,0) and larger electron pockets at M points, in general agreement with full-potential linearized plane wave calculation\cite{CLiuPRL2008Ba122}.  Upon K doping, the hole pocket expands and the electron pocket becomes smaller and the bottom of the electron Fermi surface almost approaches $E_F$. Such an evolution of the Fermi surface is consistent with hole doping within a rigid-band shift model. Furthermore, Sato {\it et al.} have performed ARPES measurements of the end-member compound KFe$_2$As$_2$, in which holes are heavily doped\cite{SatoARPESK122}. Although hole pockets around the $\Gamma$ point are qualitatively similar to those of optimally doped Ba$_{0.4}$K$_{0.6}$Fe$_2$As$_2$ with $T_c$ =37 K, the electron pockets around the M point are completely absent owing to excess hole doping. They suggest that the interband scattering via the antiferromagnetic wave vector is suppressed by the absence of an electronic pocket around the M point, and is essentially related to $T_c$ in the overdoped region.

\subsubsection{Normal-state magnetic properties in LaFeAs(O$_{1-x}$F$_x$)}
Systematic bulk susceptibility $\chi_{\rm bulk}$ measurements in LaFeAs(O$_{1-x}$F$_x$) were performed by Kohama {\it et al.}\cite{KohamaFeAsPRB2008} and Klingeler {\it et al}\cite{KlingelerLaFeAsOF}. The $\chi_{\rm bulk}$ of LaFeAs(O$_{1-x}$F$_x$) is independent of local-moment effects by rare-earth elements, and is affected only by the iron $3d$ electrons. In both reports, the $\chi_{\rm bulk}$ of LaFeAsO shows a gradual decrease from room temperature and exhibits a small drop at $T_S$ where the structural transition occurs. Below approximately 50 K, $\chi_{\rm bulk}$ shows an upturn behavior, which would be ascribed to an impurity contribution. Antiferromagnetic anomaly at $T_N$ was visible in the temperature derivative of $\chi_{\rm bulk}$ as shown in Fig. \ref{Kondrat}, which corresponds to a magnetic specific heat. 

The temperature dependences of the $\chi_{\rm bulk}$ of the F-doped compounds are different between the above two reports. The $\chi_{\rm bulk}$ of the F-doped sample reported by Kohama {\it et al.}\cite{KohamaFeAsPRB2008} shows an increasing behavior with decreasing temperature, but that reported by Klingeler {\it et al.}\cite{KlingelerLaFeAsOF} shows a gradual decrease with decreasing temperature as in the undoped LaFeAsO. Since the temperature dependence of the $^{75}$As-Knight shift is scaled to the latter behavior\cite{ImaiPRB2008, Grafe, ImaiProceedings}, the former temperature dependence is considered to be significantly affected by impurity phases. Klingeler {\it et al.} also reported that the temperature dependences of $\chi_{\rm bulk}$ well above $T_N$ and $T_c$ are barely unchanged in a wide doping region, although the ground state changes from an orthorhombic antiferromagnetic poor metal to a tetragonal nonmagnetic superconductor upon F doping. 

The dynamical susceptibility arising from the Fe-$3d$ spin dynamics has been investigated by measuring the nuclear spin-lattice relaxation rate $1/T_1$ with a nuclear-magnetic-resonance (NMR) technique\cite{NakaiJPSJ2008,ImaiPRB2008, Grafe, MukudaJPSJ08, TerasakiFeNMR, ImaiProceedings}. $1/T_1T$ is related to the $q$-integral of the low-energy component of spin fluctuations, that is,
\begin{equation*}
\frac{1}{T_1T} \sim \sum_{q} |A(q)|^2 \frac{\chi''(q,\omega)}{\omega} , 
\end{equation*}                  
where $A(q)$ and $\chi''(q,\omega)$ are the $q$-dependent hyperfine coupling constant at the observed site arising from the Fe-$3d$ spin and the dynamical susceptibility, respectively, and the sum is taken over the 1st Brillouin zone. 

\begin{figure}[tbp]
\begin{center}
\includegraphics[width=7cm]{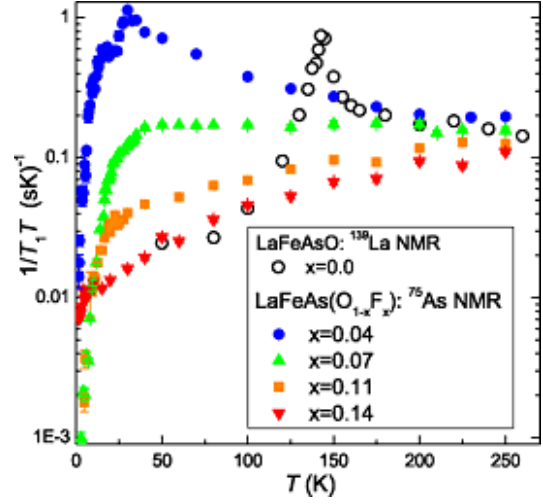}
\end{center}
\caption{(Color online) Temperature dependence of $1/T_1T$ of $^{75}$As in LaFeAs(O$_{1-x}$F$_x$) with various F concentrations. $1/T_1T$ of $^{139}$La in LaFeASO is normalized to that of $^{75}$As using the relation $^{139}(1/T_1) \propto ^{75}(1/T_1)$\cite{NakaiPRL2009}.}
\label{NakaiT1T}
\end{figure}
Systematic measurements of $1/T_1T$ in LaFeAs(O$_{1-x}$F$_x$) were performed by Nakai {\it et al}. Figure \ref{NakaiT1T} shows the temperature dependence of the $1/T_1T$ of LaFeAs(O$_{1-x}$F$_x$)\cite{NakaiPRL2009}. In this figure, the $1/T_1T$ of $^{139}$La in undoped LaFeAsO is normalized to that of $^{75}$As using the ratio of hyperfine coupling constants. The $1/T_1T$ of undoped LaFeAsO shows a clear peak at $T_{N} \sim 142$ K, which is in contrast to the tiny anomaly in $\chi_{\rm bulk}$. These are characteristic features of an itinerant antiferromagnet, in which the ordered $q$-vector is far from $q = 0$. As shown in Fig.~\ref{NakaiResT1T}, the magnetic fluctuations are related to the structural phase transition occurring below $T_S$.

The magnetic fluctuations are significantly suppressed upon F doping. In the F4\%-doped sample, $1/T_1T$ shows a kink at approximately 30 K, which is ascribed to a magnetic anomaly, and then exhibits a sharp decrease owing to the superconducting transition at 16 K. It was revealed that two anomalies are observed in the F4\%-doped sample. In the F7\%-doped sample, $1/T_1T$ is nearly constant from room temperature, and starts to decrease at 40 K, which is far above $T_c \sim$ 20 K. The decrease in $1/T_1T$ above $T_c$ becomes significant upon further F doping, and the $1/T_1T$ of the F11\%-doped sample shows a continuous decrease from room temperature. The decrease in $1/T_1T$ far above $T_c$ is reminiscent of the pseudogap behavior observed in underdoped cuprates\cite{TakigawaPRB1991}. This pseudogap behavior was first pointed out by Ahilan {\it et al.} from their F-NMR measurement on LaFeAs(O$_{0.89}$F$_{0.11}$)\cite{ImaiPRB2008}.
By using the empirical formula 
\begin{equation*}
\frac{1}{T_1T} = a + b\exp{\left(-\frac{\Delta_{\rm PG}}{T}\right)}, 
\end{equation*}
the magnitude of the pseudogap is estimated to be $\Delta_{\rm PG} = 172 \pm 12$ K with $a = 0.04$ (s$^{-1}$K$^{-1}$) and $b = 0.19$ (s$^{-1}$K$^{-1}$) for the F11\%-doped sample and $\Delta_{\rm PG} = 165 \pm 15$ K with $a = 0.012$ (s$^{-1}$K$^{-1}$) and $b = 0.18$ (s$^{-1}$K$^{-1}$) for the F14\%-doped sample. The estimated pseudogap energies $\Delta_{\rm PG}$ of the F11\%- and F14\%-doped samples are almost the same within experimental errors\cite{NakaiPRL2009, ImaiProceedings}. One of the significant differences between the iron oxypnictides and the cuprates is that the pseudogap behavior becomes more pronounced in the heavily F-doped region in LaFeAs(O$_{1-x}$F$_x$), whereas it does in the underdoped region in the cuprate superconductors. 

\begin{figure}[tbp]
\begin{center}
\includegraphics[width=7.5cm]{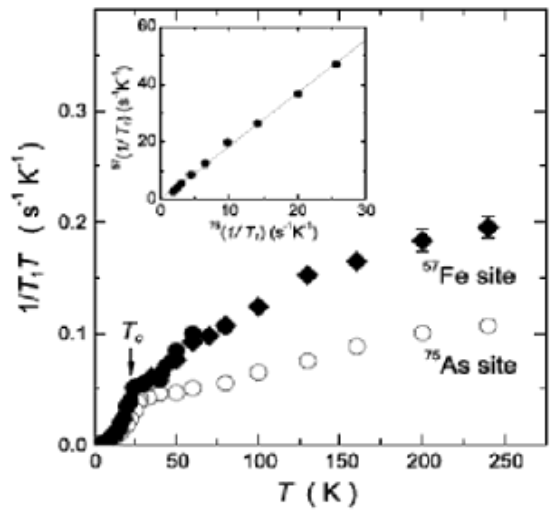}
\end{center}
\caption{Temperature dependence of $1/T_1T$ of $^{57}$Fe, together with $1/T_1T$ of $^{75}$As in LaFeAsO$_{0.6}$. The inset shows a plot of $^{57}(1/T_1T)$ against $^{75}(1/T_1T)$ as an implicit parameter $T$ between 30 and 240 K\cite{TerasakiFeNMR}. Figure reprinted from N. Terasaki {\it et al.}: J. Phys. Soc. Jpn {\bf 78} (2009) 013701. Copyright 2009 by the Japanese Physical Society.}
\label{TerasakiFeNMRT1T}
\end{figure}
In addition, Terasaki {\it et al.} performed $^{57}$Fe and $^{75}$As-NMR measurements in LaFeAsO$_{0.7}$, and reported that the $1/T_1T$ of $^{57}$Fe shows a similar temperature dependence to that of $^{75}$As, as shown in Fig.~\ref{TerasakiFeNMRT1T}\cite{TerasakiFeNMR}. This indicates that the magnitudes of pseudogap at the Fe and As sites are almost identical, and that low-energy spin fluctuations are suppressed over the whole $q$-space with decreasing temperature, since the $1/T_1T$ of Fe is determined by the spin fluctuations over the entire $q$-space. These systematic $^{75}$As-NMR measurements of LaFeAs(O$_{1-x}$F$_x$) and $^{57}$Fe-NMR in LaFeAsO$_{1-\delta}$ suggest that the psuedogap cannot be ascribed to antiferromagnetic fluctuations. Recently, it has been proposed that the pseudogap originates from the characteristic band structure near $E_F$\cite{Ikeda}. A hole Fermi surface around $\Gamma'(\pi,\pi)$ in the {\it unfolded} BZ, existing just near $E_F$, sinks down with electron doping. Then, thermally excited electrons can contribute to $N(E_F)$ at higher temperatures, resulting in the observed pseudogap behavior in the electron-doped region. This scenario is consistent with the experimental data. For further clarification of the origin of the pseudogap, particularly $1/T_1$ measurements in the hole-doped region are desired.      

\subsubsection{Normal-state magnetic properties in Ba(Fe$_{1-x}$Co$_x$)$_2$As$_2$}
\begin{figure}[tbp]
\begin{center}
\includegraphics[width=8cm]{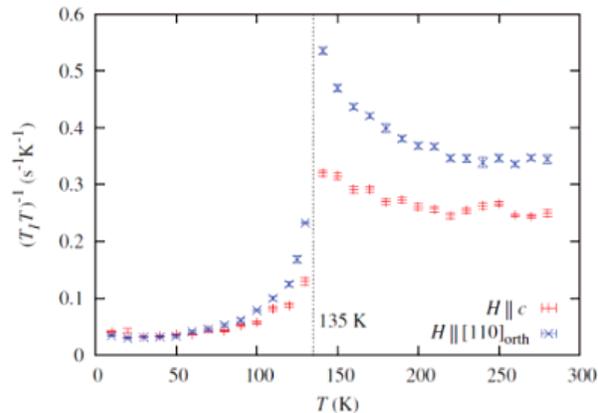}
\end{center}
\caption{(Color online) $1/T_1T$ of $^{75}$As in BaFe$_2$As$_2$ is plotted as a function of temperature for two field orientations. The upturn of $1/T_1T$ in the paramagnetic state was observed only for $H \perp c$, indicative of the anisotropic development of the stripe-type antiferromagnetic spin fluctuations perpendicular to both the $c$-axis and the stripe\cite{KitagawaBaFe2As2}. Figure reprinted from K. Kitagawa {\it et al.}: J. Phys. Soc. Jpn {\bf 77} (2008) 114709. Copyright 2008 by the Japanese Physical Society.}
\label{KitagawaT1T}
\end{figure}

NMR studies of single-crystal and poly-crystal BaFe$_2$As$_2$ with the ``122'' structure were performed by several groups\cite{FukazawaJPSJ08, KitagawaBaFe2As2, BaekBaFe2As2}. In contrast with LaFeAsO, Kitagawa {\it et al.} reported a first-order-type antiferromagnetic transition from the splitting of $^{75}$As NMR lines in BaFe$_2$As$_2$, which is accompanied by the simultaneous structural transition as evidenced by a sudden large change in the electric field gradient tensor at the As site\cite{KitagawaBaFe2As2}. From the analyses of the hyperfine field at the As site, a spin structure in the antiferromagnetic state was microscopically determined: a stripe structure with the antiferromagnetic moments directed perpendicular to the stripe is formed. This stripe structure is fully consistent with the results of neutron-diffraction experiments. 
The anisotropic temperature dependence of $1/T_1T$ was observed in the paramagnetic state, which is shown in Fig.~\ref{KitagawaT1T}. $1/T_1T$ for $H \perp c$ increases with decreasing temperature, while $1/T_1T$ for $H \parallel c$ is nearly constant down to 175 K. This anisotropy in $1/T_1T$ is explained in terms of the fluctuations of the hyperfine field along the $c$-direction, which can be generated only from the above stripe-type spin fluctuations along the $a$-direction.

\begin{figure}[tbp]
\begin{center}
\includegraphics[width=7.5cm]{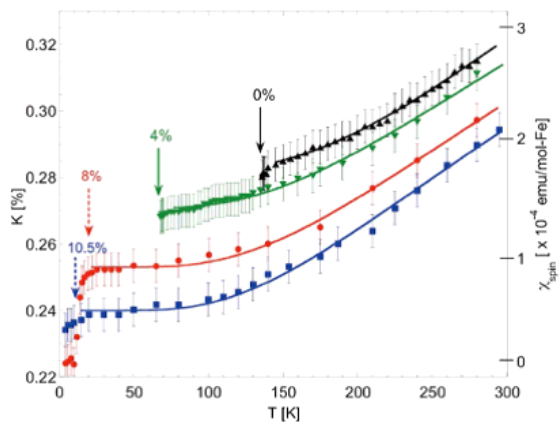}
\end{center}
\caption{(Color online) Temperature-dependent of the gravity of the $^{75}$As-NMR Knight shift $K$ measured along the crystal $c$-axis for Ba(Fe$_{1-x}$Co$_x$)$_2$As$_2$. $K$ = 0.224\% corresponds to the background arising from the chemical shift $K_{\rm chem.}$. The conversion to spin susceptibility $\chi_{spin}$ based on the eq.(b) is shown on the right vertical axis. Vertical solid and dashed arrows denote $T_{\rm SDW}$ and $T_c$, respectively. Solid curves are fits to the activation formula $1/T1T = a +b \exp{(-\frac{\Delta_{\rm PG}}{T})}$ with $\Delta_{\rm PG}$ = 711 K ($x = 0$), 570 K ($x = 0.04$), 520 K ($x = 0.08$), and 490 K ($x = 0.105$)\cite{NingBaFeCoAs}. Figure reprinted from F. L. Ning {\it et al.}: J. Phys. Soc. Jpn {\bf 78} (2009) 013711. Copyright 2009 by the Japanese Physical Society.}
\label{NingBaFeCoAs}
\end{figure}

The evolution of normal-state magnetic properties with electron doping was systematically studied by Ning {\it et al.} from the $^{75}$As-NMR measurements in single-crystal Ba(Fe$_{1-x}$Co$_x$)$_2$As$_2$, where superconductivity emerges upon doping Co at the Fe site\cite{NingBaFe1.8Co0.2As2, NingBaFeCoAs}.
The intrinsic spin susceptibility and low-energy spin dynamics were investigated by measuring the Knight shift and $1/T_1T$ at $x =$ 0.00 ($T_N$ = 135 K), 0.04 ($T_N$ = 66 K), 0.08 ($T_c$ = 22 K), and 0.105 ($T_c$ = 15 K). Figure \ref{NingBaFeCoAs} shows the $x$ and temperature dependences of the Knight shift. In the figure, the spin susceptibility $\chi_{\rm spin}$ is shown on the right vertical axis, which was estimated using,
\begin{eqnarray*}
K &=& K_{\rm spin} + K_{\rm orb} \hspace{2cm} {\rm (a)} , \\
K_{\rm spin} &=& \frac{A_{\rm hf}}{N_A\mu_{B}}\chi_{\rm spin} \hspace{2.5cm} {\rm. (b)}
\end{eqnarray*}
Here, $K_{\rm spin}$ ($K_{\rm orb}$) is the spin (orbital) part of the Knight shift, and $A_{\rm hf}$ is the hyperfine coupling constant.
It was shown that $\chi_{\rm spin}$ decreases with decreasing temperature and is suppressed by electron doping. Ning {\it et al.} pointed out that the some energy scale such as the effective spin-spin exchange interaction or pseudogap increases with electron doping in Ba(Fe$_{1-x}$Co$_x$)$_2$As$_2$. This tendency seems to be consistent with that in LaFeAs(O$_{1-x}$F$_x$), but is opposite to that in high-$T_c$ cuprates, where such an energy scale is suppressed by hole doping\cite{NingBaFeCoAs}. 
\begin{figure}[tbp]
\begin{center}
\includegraphics[width=7.5cm]{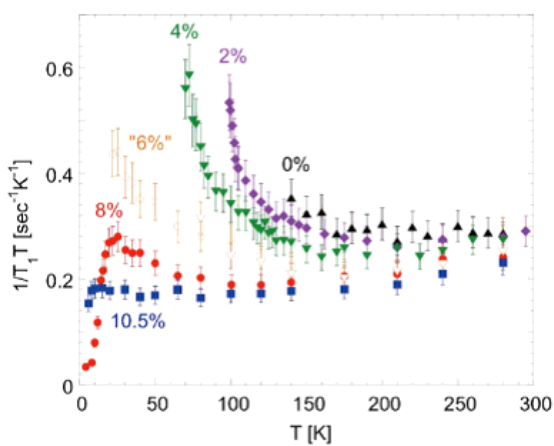}
\end{center}
\caption{(Color online) Filled symbols show $1/T_1T$ for $H \parallel c$ measured at the peak of the As NMR lineshapes for Ba(Fe$_{1-x}$Co$_x$)$_2$As$_2$ with 0, 2, 4, 8, and 10.5\% Co dopings. Orange open circles (triangles) labeled as ``6\%'' were deduced from the half-intensity position of the FFT lineshape of the 8\% (4\%) doped sample on the higher (lower)-frequency side\cite{NingBaFe1.8Co0.2As2,NingBaFeCoAs}. Figure reprinted from F. L. Ning {\it et al.}: J. Phys. Soc. Jpn {\bf 78} (2009) 013711. Copyright 2009 by the Japanese Physical Society.}
\label{T1TNingBaFeCoAs}
\end{figure}
Figure \ref{T1TNingBaFeCoAs} shows the Co concentration $x$ and temperature dependences of $1/T_1T$ in Ba(Fe$_{1-x}$Co$_x$)$_2$As$_2$\cite{NingBaFe1.8Co0.2As2,NingBaFeCoAs}. Although $x =$ 0.02 and 0.04 samples show a divergent behavior of $1/T_1T$ due to critical slowing down of spin fluctuations toward $T_N$, $1/T_1T$ at $x$ = 0.105 levels off after a monotonic decrease with decreasing temperature,  and exhibits no evidence for enhancement owing to low-energy antiferromagnetic fluctuations. Considering that $\chi_{\rm spin}$ is also suppressed below 300 K for all concentrations, it is considered that low-energy spin fluctuations are suppressed with decreasing $T$ except near $T_N$, i.e. Ba(Fe$_{1-x}$Co$_x$)$_2$As$_2$ exhibit pseudogap behavior for broad concentrations as in LaFeAs(O$_{1-x}$F$_x$).

\subsection{Superconducting properties}
Superconducting properties have been investigated through various experiments. Here, some key experimental results are reviewed.

\subsubsection{Upper critical field $H_{c2}$}
\begin{figure}[tbp]
\begin{center}
\includegraphics[width=7cm]{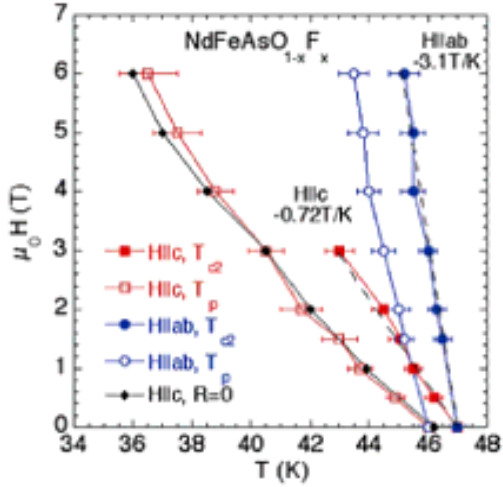}
\end{center}
\caption{(Color online) Phase diagram of NdFeAsO$_{1-x}$F$_x$ as determined from the field dependence of the peak in the specific-heat measurements and of $T_c$ determined through an entropy conserving construction\cite{WelpPRB2008}. Figure reprinted from U. Welp {\it et al.}: Phys. Rev. B {\bf 78} (2008) 140510(R). Copyright 2008 by the American Physical Society.}
\label{Hc2NdFeAsOF}
\end{figure}

In the early stage, Hunte {\it et al.} reported resistive measurements of polycrystalline LaFeAsO$_{0.89}$F$_{0.11}$ at high magnetic fields of up to 45 T, that show a remarkable increase in the upper critical field $H_{c2}$ compared with values expected from the slopes d$H_{c2}$/d$T \sim -2$ T/K near $T_c$ at low temperatures\cite{HunteNature2008}. Since this enhancement behavior is similar to that observed in dirty MgB$_2$ films, they suggested that superconductivity in LaFeAsO$_{0.89}$F$_{0.11}$ results from two bands (i.e. a nearly two-dimensional diffusive electron band and a more isotropic heavy hole band), and that the $H_{c2}$ enhancement behavior can be interpreted by a two-gap scenario. However, to confirm these issues, it seems that single-crystal measurements are highly desired.        
Welp {\it et al.} reported the upper critical field $H_{c2}$ of single-crystal NdFeAs(O$_{1-x}$F$_x$) with $T_c \sim 47$ K from specific-heat measurements\cite{WelpPRB2008}. Figure \ref{Hc2NdFeAsOF} shows a phase diagram of NdFeAs(O$_{1-x}$F$_x$) as determined from the field dependence of superconducting anomalies in specific-heat and resistivity measurements. The average slopes of $H_{c2}$ for the $c$- and $ab$-axes are respectively $dH_{c2}^{c}/dT=-0.72$ T/K and $dH_{c2}^{ab}/dT=-3.1$ T/K, resulting in a modest superconducting anisotropy of $\Gamma \equiv H_{c2}^{ab}/H_{c2}^{c} \simeq 4.3$\cite{WelpPRB2008}. 
Using the ordinary single-band Ginzburg-Landau relations for the upper critical field, i.e, $H_{c2}^{c}=\phi_0/2\pi\xi_{ab}^2$ and $H_{c2}^{ab}=\phi_0/2\pi\xi_{ab}\xi_{c}$, and the single-band Werthamer-Helfand-Hohenberg (WHH) expression\cite{WHH} $H_{c2}(0)=-0.69T_c(dH_{c2}/dT)|_{T_c}$, $H_{c2}^{c}(0)\simeq 23$ T, $H_{c2}^{ab}(0)\simeq 100$ T, $\xi_c(0)\simeq0.9$ nm, and $\xi_{ab}(0)\simeq3.7$ nm are evaluated.
\begin{figure}[tbp]
\begin{center}
\includegraphics[width=7cm]{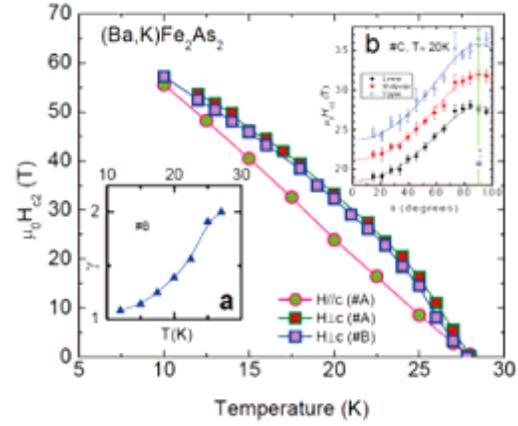}
\end{center}
\caption{(Color online) The main figure shows $H_{c2}$ vs temperature for magnetic fields parallel (circles) and perpendicular (squares) to the $c$-axis, in which the critical fields $H_{c2}$(T) are determined from the midpoint of the sharp resistive superconducting transitions. The two samples (\#A and \#B) behave nearly identically. Inset {\bf a}: The anisotropy parameter $\Gamma = H_{c2}^{H\perp c}/H_{c2}^{H \| c}$ plotted as a function of temperature. Inset {\bf b}: The upper critical field $\mu_{0}H_{c2}$ (sample \#C, $T_c \sim$ 28 K), derived at the upper, middle point, and lower limits of the main resistive transition, is plotted as a function of the tilt angle $\theta$ at $T = 20$ K, where $\theta$ is the angle between the applied magnetic field ($H$) and the crystallographic $c$-axis. The angular dependence of $\mu_{0}H_{c2}$ can be well scaled by $H_{c2}(\theta) =H_{c2}[\cos^2{\theta}+\Gamma^{-2}\sin^2{\theta}]^{-0.5}$  with $\Gamma =1.5 \pm 0.1$, which is very close to the value ($1.4 \pm 0.1$) derived from the inset {\bf a}\cite{Yuan}. Figure reprinted from H. Q. Yuan {\it et al.}: Nature. {\bf 457} (2009) 565. Copyright 2009 by Macmillian Publishers Limited.}
\label{Hc2(BaK)Fe2As2}
\end{figure}

$H_{c2}$ of single-crystal (BaK)Fe$_2$As$_2$ was investigated by Yuan {\it et al}\cite{Yuan}. 
Figure~\ref{Hc2(BaK)Fe2As2} displays the $H_{c2}$ determined at the resistive midpoint vs temperature for two different single-crystalline samples of (Ba,K)Fe$_2$As$_2$ (\#A and \#B) up to 60 T. Although their residual resistivities are different ($\rho_0(\#A)=0.12$ m$\Omega$cm and $\rho_0(\#B)=0.077$ m$\Omega$cm), the two samples have almost identical $T_c(\simeq28$ K) and $H_{c2}$ behaviors. This suggests that superconductivity in (Ba,K)Fe$_2$As$_2$ is not so sensitive to disorder or impurity scattering, in contrast to some other unconventional superconductors.
The most remarkable aspect in Fig.~\ref{Hc2(BaK)Fe2As2} is the fact that the $H_{c2}$ of (Ba,K)Fe$_2$As$_2$ extrapolates to a similar value of 70 T as $T \to 0$, irrespective of directions for applied magnetic fields.  This is in great contrast to the behavior of other quasi-2D superconductors such as organic superconductors or cuprates, where the in-plane critical fields are much larger than fields applied perpendicular to the planes.
A superconducting anisotropy $\Gamma \equiv H_{c2}^{ab}/H_{c2}^{c}$ at about $T_c$ was also investigated in Ba$_{0.6}$K$_{0.4}$Fe$_2$As$_2$. Unexpectedly, a small $\Gamma \sim2$ was reported at about $T_c$\cite{Z.S.WangPRB2008}, which is lower than that of NdFeAs(O$_{1-x}$F$_x$)\cite{WelpPRB2008}.
The small anisotropic $\Gamma$ might be understood in terms of the 3D-like Fermi surfaces of (Ba,K)Fe$_2$As$_2$: larger corrugations in the Fermi surfaces are sufficient to permit closed orbits at all field orientations; hence, orbital limiting effects will persist at all field angles, leading to the rather isotropic upper critical field observed.

The superconducting parameters for single crystals of NdFeAs(O$_{1-x}$F$_{x}$) and (Ba$_{1-x}$K$_x$)Fe$_2$As$_2$ are summarized in Table IV.

\begin{table*}[t]
\caption
{Superconducting parameters in single crystals of NdFeAs(O$_{1-x}$F$_x$) and (Ba$_{1-x}$K$_x$)Fe$_2$As$_2$. To estimate the superconducting coherence length $\xi$, ordinary single-band Ginzburg-Landau relations are applied.}
\begin{tabular}{cc|cccccccc} \hline \hline
sample&Ref. & $T_c$ [K] & $H_{\rm apply}$ [T] & $\frac{dH_{c2}^{c}}{dT}$ [K/T] & $\frac{dH_{c2}^{ab}}{dT}$ [K/T] & $\xi_{c}$ [nm] & $\xi_{ab} [nm] $ & $H_{c2}^c$ [T] & $H_{c2}^{ab}$ [T] \\ \hline 
NdFeAs(O$_{0.82}$F$_{0.18}$)&\cite{WelpPRB2008} & 47 & $< 6$ & $-0.72$ & $-3.1$ & 0.9 & 3.7 & 23 & 100  \\
NdFeAs(O$_{0.82}$F$_{0.18}$)&\cite{JinAPL2008}& 50 & $< 9$ & $-2.1$  & $-9$   & &  & 62-70 & 304  \\
NdFeAs(O$_{0.7}$F$_{0.3}$)&\cite{JaroszynskiPRB08} & 50 & $< 60$ & $\sim -0.8$  & $\sim -10$ &0.26 & 2.3 & &  \\
(Ba$_{0.6}$K$_{0.4}$)Fe$_2$As$_2$&\cite{Z.S.WangPRB2008}& 36.5 & $< 9$ & $-5.49$ & $-9.35$ &  &  & 138 & 235  \\
(Ba$_{0.55}$K$_{0.45}$)Fe$_2$As$_2$&\cite{Altarawneh}& 32 & $ 60 $ &  &  & $\sim 2.9$ & $\sim 3.4$ &$>70$&$>70$   \\
(Ba$_{1-x}$K$_{x}$)Fe$_2$As$_2$&\cite{Yuan}& 28.5 & $< 60$ & $-2.9$ & $-5.4$ &  &  & $\sim 100$ &$\sim70$  \\
\hline \hline

\end{tabular}
\end{table*}

\subsubsection{ARPES measurements in the superconducting state}
ARPES in the superconducting state is an excellent method for investigating the symmetry of the order parameter, since ARPES can directly measure the $k$ dependence of superconducting gap. At present, since available high-quality single crystals are (BaK)Fe$_2$As$_2$ and NdFeAs(O$_{1-x}$F$_x$), ARPES measurements have been performed in these compounds\cite{KondoPRL2008, LiuNdFeAsOF, HDingBaKFe2As2, ZhaoBaKFe2As2, CLiuPRL2008Ba122, HLiuPRB2008ARPES, YangARPESparentBa122, YZhangARPESSrK122, DingARPESBaK122}.

\begin{figure}[tbp]
\begin{center}
\includegraphics[width=7cm]{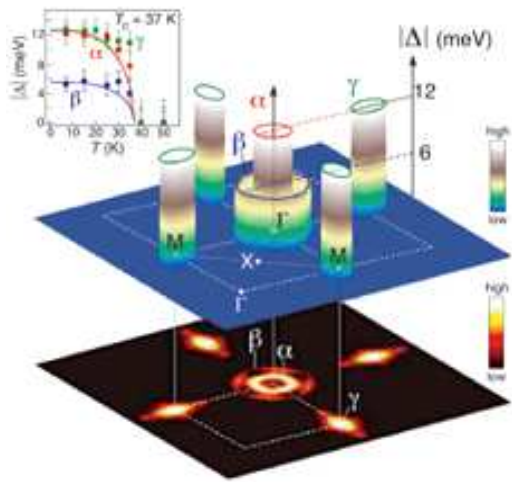}
\end{center}
\caption{(Color online) Three-dimensional plot of the superconducting-gap size ($\Delta$) measured at 15 K on the three observed Fermi-surface sheets shown at the bottom as an intensity plot\cite{HDingBaKFe2As2}. The inset shows the temperature dependence of these three gaps. Figure reprinted from H. Ding {\it et al.}: EuroPhys. Lett. {\bf 83} (2008) 47001. Copyright 2008 by EDP Sciences.}
\label{ARPESDing}
\end{figure}
 Ding {\it et al.} have performed ARPES measurements on the superconducting Ba$_{0.6}$K$_{0.4}$Fe$_2$As$_2$ ($T_c = 37$ K), and the results are summarized in Fig.~\ref{ARPESDing}\cite{HDingBaKFe2As2}. Two superconducting gaps with different values were observed : a larger gap ($\Delta\sim$12meV) on small hole-like and electron-like Fermi-surface sheets, and a small gap ($\Delta\sim$6meV) on the larger hole-like Fermi-surface sheet, which are shown in the main panel. Both superconducting gaps are nearly isotropic without nodes, and close simultaneously at the bulk $T_c$, as shown in the inset. Since the large superconducting gap ascribed to strong pairing interactions is observed in the two small hole and electron Fermi surfaces connected by the ($\pi$,0) spin-density-wave vector, Ding {\it et al.} suggest that the pairing mechanism originates from the inter-band interactions between these two nested Fermi-surface sheets. Similar experimental results have been reported from different groups\cite{ZhaoBaKFe2As2, CLiuPRL2008Ba122, HLiuPRB2008ARPES, YangARPESparentBa122, YZhangARPESSrK122, DingARPESBaK122}. 

In the ``1111'' system, ARPES measurements have been performed in the superconducting NdFeAsO$_{0.9}$F$_{0.1}$\cite{KondoPRL2008, LiuNdFeAsOF}. It was found that the hole-like Fermi surface around the $\Gamma$ point is fully gapped below $T_c$ and that the superconducting gap is $15 \pm 1.5$ meV at an anisotropy smaller than 20\% of the SC gap. The superconducting gap at the electron-like Fermi surface is not reported at present. 
These ARPES measurements of Ba$_{1-x}$K$_x$Fe$_2$As$_2$ and NdFeAsO$_{0.9}$F$_{0.1}$ strongly suggest that 1) the superconducting gap is nodeless, which excludes the $p$-wave and $d$-wave pairing states with nodes of the gap on the Fermi surfaces but consistent with $s$-wave pairing, and that 2) the magnitude of superconducting gap is orbital-dependent.

\subsubsection{Magnetic penetration-depth measurements}
\begin{figure}[tbp]
\begin{center}
\includegraphics[width=6.5cm]{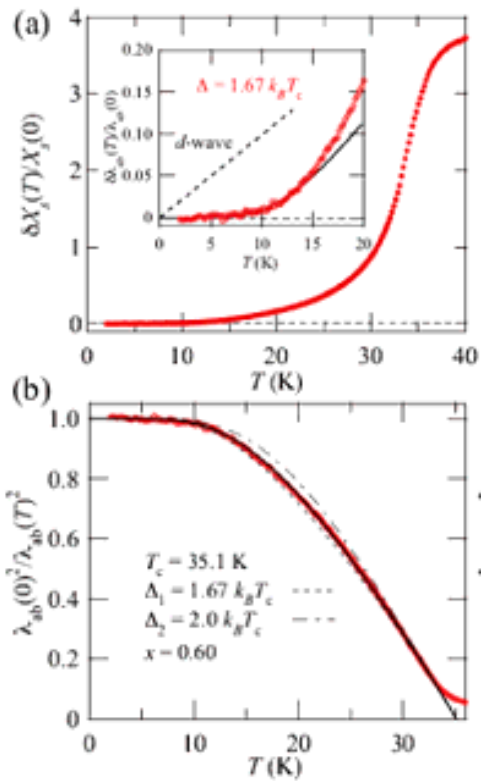}
\end{center}
\caption{(Color online) (a) Temperature dependence of $\delta X_s(T)/X_s(0)$, which corresponds to $\delta\lambda_{ab}(T)/\delta\lambda_{ab}(0)$ in the superconducting state. The inset is an expanded view at low temperatures. The dashed line represents a $T$-linear dependence expected in clean $d$-wave superconductors with line nodes. The solid line is a low-$T$ fit to $\frac{\delta\lambda_{ab}(T)}{\lambda_{ab}(0)} \sim \sqrt{\frac{\pi\Delta}{2k_BT}}\exp{\left(-\frac{\Delta}{k_BT_c}\right)}$. (b) Temperature dependence of the superfluid density $\lambda_{ab}^{2}(0)/\lambda_{ab}^{2}(T)$. The solid line is the best fit results to the two-gap model, and the dashed and dashed-dotted lines are the single gap results for $\Delta_1$ and $\Delta_2$, respectively. $T_c$ is defined by the temperature at which the superfulid density becomes zero\cite{HashimotoPrFeAsO}. Figures reprinted from K. Hashimoto {\it et al.}: Phys. Rev. Lett {\bf 102} (2009) 017002. Copyright 2009 by the American Physical Society.}
\label{PDHashimoto}
\end{figure}

Nodeless superconducting gap was also suggested from magnetic penetration-depth measurements. Hashimoto {\it et al.} measured the in-plane microwave penetration depth $\lambda_{ab}$ in underdoped single crystals of PrFeAsO$_{1-\delta}$ ($T_c\sim 35$ K), and reported that $\lambda_{ab}(T)$ shows a flat temperature dependence at low temperatures, as shown in Fig.~\ref{PDHashimoto}\cite{HashimotoPrFeAsO}. This result is incompatible with the presence of nodes and suggests a nodeless superconducting gap with $\Delta_{\rm min}/k_BT_c >$ 1.6 all over Fermi surface. However, it was reported that the whole temperature dependence of the penetration depth cannot be reproduced by the single superconducting gap estimated from the low-temperature behavior, but by two different isotropic gaps with $\Delta/k_BT_c$ = 1.67 and 2.0. This simple model was successfully used for the two-gap $s$-wave superconductor MgB$_2$, and suggests the presence of two different gaps also in PrFeAsO$_{1-\delta}$. A similar two-nodeless-gap model is also applied in SmFeAsO$_{1-x}$F$_x$ ($T_c = 45$ K)\cite{MaloneSmFeAsOF}, but the result suggesting a single fully gapped order parameter with a small anisotropy $\Delta_0^{\rm max}/\Delta_0^{\rm min} \sim 1.2$ was reported in single crystals of NdFeAsO$_{0.9}$F$_{0.1}$\cite{MartinNdFeAsOF}.

High-sensitive microwave measurements of the in-plane penetration depth $\lambda_{ab}$ were performed in several single crystals of the hole-doped superconductor Ba$_{1-x}$K$_x$Fe$_2$As$_2$\cite{HashimotoBaKFe2As2}. It was reported that the temperature dependence of $\lambda_{ab}$ is sensitive to degrees of disorder inherent in the crystals. The exponential temperature dependence consistent with a fully opened gap was observed at the penetration depth for the cleanest crystal at low temperatures, which contradicts the $T$-linear dependence expected in clean $d$-wave superconductors with line nodes. The overall temperature dependence of the penetration depth cannot be fully reproduced by a single gap calculation, but can be successfully explained by a two-gap model with $\Delta_1/k_BT_c$ = 1.17 and $\Delta_2/k_BT_c$ = 2.40. In addition, a linear relation between the low-temperature scattering rate $1/\tau(T)$ and the density of quasiparticles $n(T)$ was observed, which is also in contrast to that in the case of $d$-wave cuprate superconductors with nodes in the gap, but is consistent with the case of $s$-wave superconductors without nodes, since $1/\tau(T)$ and $n(T)$ in an $s$-wave superconductor are both expected to follow the exponential dependence at low temperatures, resulting in a linear relation between these two quantities. A similar two-isotropic-gap state was also suggested from penetration-depth measurements with the $\mu$SR method\cite{Hiraishi(BK)Fe2As2muSR}. 

\subsubsection{NMR in the superconducting state}
\begin{figure}[tbp]
\begin{center}
\includegraphics[width=6.5cm]{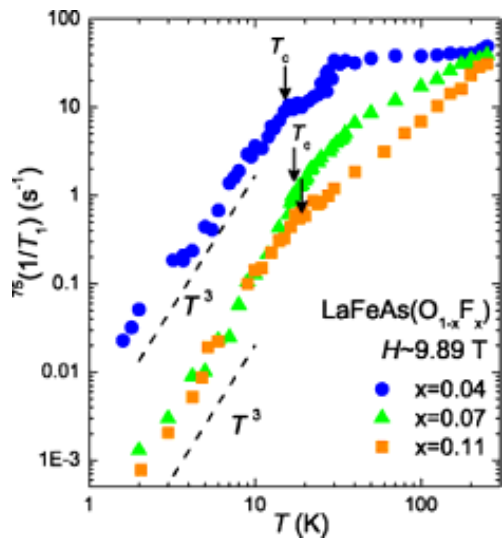}
\end{center}
\caption{(Color online) Temperature dependence of $1/T_1$ of $^{75}$As, measured at the peak corresponds to $H \parallel$ the $ab$-plane\cite{NakaiPRL2009}.}
\label{T1Nakai}
\end{figure}
\begin{figure}[tbp]
\begin{center}
\includegraphics[width=6.5cm]{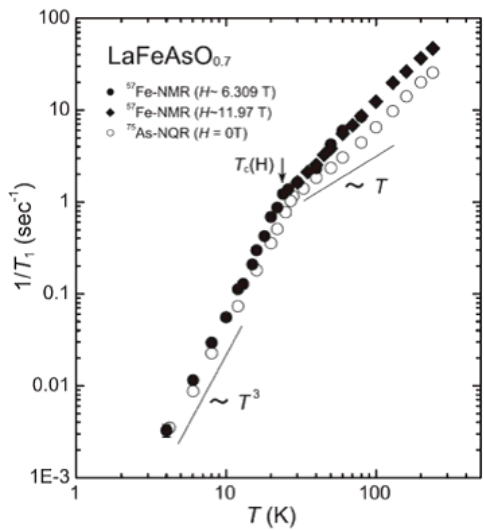}
\end{center}
\caption{Temperature dependences of $1/T_1$ of $^{57}$Fe and $^{75}$As in LaFeAsO$_{0.6}$ with $T_c$ = 28 K\cite{TerasakiFeNMR}. Figure reprinted from N. Terasaki {\it et al.}: J. Phys. Soc. Jpn {\bf 78} (2009) 013701. Copyright 2009 by the Japanese Physical Society}
\label{T1Terasaki}
\end{figure}

Up to now, several NMR groups have reported the temperature dependence of the nuclear spin-lattice relaxation rate $1/T_1$ in the superconducting state, which is related to the superconducting-gap structure\cite{NakaiJPSJ2008, NakaiPRL2009, ImaiPRB2008, Grafe, MukudaJPSJ08, TerasakiFeNMR, NingBaFe1.8Co0.2As2, NingBaFeCoAs, ImaiProceedings, MatanoEPL2008, SKawasakiTwoGap, GrafeNJP}. Nakai {\it et al.} reported the $T$-dependences of $1/T_1$ of $^{75}$As for $x=0.04$, 0.07 and 0.11, which were measured in $H \sim 9.89$ T parallel to the $ab$-plane\cite{NakaiJPSJ2008, NakaiPRL2009}. The experimental results of $1/T_1$ are shown in Fig.~\ref{T1Nakai}. It was found that $1/T_1$ decreases suddenly without showing a Hebel-Slichter (coherence) peak, and that a $T^3$ dependence was observed in $1/T_1$ in the superconducting state for $x$ = 0.07 and 0.11 samples as seen in Fig. \ref{T1Nakai}. In order to determine whether or not this $T^3$ dependence is modified by applying external fields, $1/T_1$ of the $x=0.11$ sample was measured in various magnetic fields. It was found that the $T^3$ behavior holds down to 4 K ($T \sim 0.2 T_c$) in the measured field range of $5.2 - 12$ T within experimental error, indicative of the $T^3$ behavior being intrinsic in the superconducting state. The absence of a coherence peak and the robust $T^3$ dependence in $1/T_1$ were also reported in LaFeAs(O$_{0.9}$F$_{0.1}$) by Grafe {\it et al.}
In addition, a similar $T^3$ behavior down to 4 K was also reported from $^{75}$As-NQR measurements in oxygen-deficient LaFeAsO$_{0.6}$, where $1/T_1$ was measured in zero field\cite{MukudaJPSJ08}. This also suggests that the $T^3$ behavior below $T_c$ is intrinsic, which is not affected by applied fields. Quite recently, Terasaki {\it et al.} have reported that the same $T^3$ temperature dependence of $1/T_1$ is observed not only at the As site but also at the Fe site where the pairing interaction is considered to arise\cite{TerasakiFeNMR}. The temperature dependence of $1/T_1$ in LaFeAsO$_{0.7}$ with $T_c = 28$ K is displayed in Fig. \ref{T1Terasaki}. It is interesting that a similar temperature dependence of $1/T_1$ was also observed in the $\alpha$- FeSe superconductor through a Se NMR\cite{KotegawaJPSJFeSe}.                   

The observed $T^3$ dependence of $1/T_1$ for LaFeAs(O$_{0.89}$F$_{0.11}$) can be reproduced using a two-dimensional line-node ($\Delta(\phi)=\Delta_0\sin{(2\phi)}$) model with $2\Delta/k_{\rm B}T_c = 4.0$\cite{NakaiJPSJ2008}. 
However, Nakai {\it et al.} pointed out that a residual density of states (DOS) suggested from the Korringa behavior at low temperatures is not observed in LaFeAs(O$_{1-x}$F$_x$)\cite{NakaiJPSJ2008, NakaiPRL2009}. In non $s$-wave superconductors with line nodes crossing the Fermi surface, the residual DOS is easily induced by a small amount of impurities and crystal imperfections, and was observed in most unconventional superconductors\cite{IshidaPRL00,CurroNature05}. In addition, it is known in anisotropic superconductors with nodes that the applied magnetic field induces the extra relaxation rate at low temperatures, which originates from the Volvik effect\cite{ZhengPRL2002}. It is pointed out that a quasiparticle state would extend outside the vortex cores with an ungapped spectrum in a $d$-wave superconductor, which gives rise to a field-induced relaxation rate. However, no such field-induced relaxation rate was observed, as seen in Figs. \ref{T1Nakai} and  \ref{T1Terasaki}. The absence of the residual DOS and the field-induced relaxation at low temperatures seems to be in contrast to that for non-$s$-wave models. It should be noted that the robust $T^3$ behavior in the superconducting state has been observed only in high-quality samples of the unconventional superconductors under zero field.  

Alternatively, recent theoretical studies have suggested that the absence of the coherence peak is consistent with the fully gapped $s_{\pm}$ state and that the $T^3$ dependence in the superconducting state can be reconciled with this $s_{\pm}$ state with an impurity effect\cite{ParkerPRB2008, ChubukovPRB2008, ParishPRB2008, Bang, NagaiNJP2008}. Here, in this $s_{\pm}$ state, the sign of the superconducting gap is reversed in the hole and electron Fermi surfaces and each superconducting gap is nodeless. In other recent theoretical studies, it was shown that a fully gapped anisotropic $s_{\pm}$-wave superconductivity with some impurity effect can account for the NMR results\cite{NagaiNJP2008}. In order to check these scenarios, further $1/T_1$ measurements using high-quality samples and iron-pnictide superconductors other than the ``1111'' system are crucial, particularly low-temperature $1/T_1$ measurements.  

Another important information from NMR measurements in the superconducting state is the symmetry of the superconducting pair, which is obtained through Knight-shift measurements, because the Knight-shift measurement is the most accurate method of measuring spin susceptibility in the superconducting state. 
Up to date, the Knight shift measurements in the superconducting state have been reported by several groups\cite{TerasakiFeNMR,NingBaFe1.8Co0.2As2,MatanoEPL2008,KawabataJPSJ08}.

\begin{figure}[tbp]
\begin{center}
\includegraphics[width=7cm]{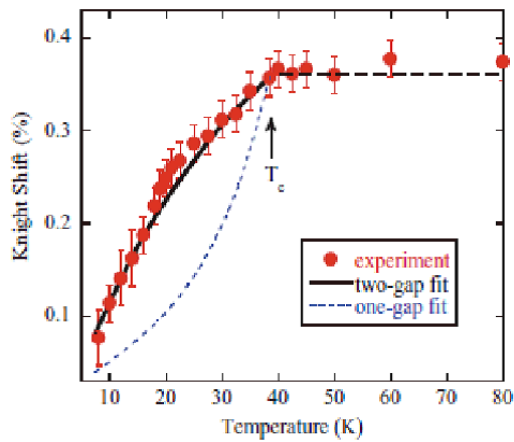}
\end{center}
\caption{(Color online) Temperature variation of the $^{75}$As Knight shift with $H \parallel ab$. The solid curve is a fit obtained using a two-gap model with $\Delta_1(T=0) = 3.5 k_BT_c$ and $\Delta_2(T=0) = 1.1 k_BT_c$. The broken curve below $T_c$ is a simulation for the larger gap alone. In both cases, $K_{\rm orb}$ was taken as 0.008\% \cite{MatanoEPL2008}. Figure reprinted from K. Matano {\it et al.}: EuroPhys. Lett. {\bf 83} (2008) 57001. Copyright 2008 by EDP Sciences.}
\label{KSMatano}
\end{figure}
Matano {\it et al.} reported the temperature dependence of the Knight shift in PrFeAsO$_{0.85}$F$_{0.15}$ with $T_c = 45$ K.  Figure \ref{KSMatano} shows the temperature dependence of the Knight shift below 80 K\cite{MatanoEPL2008}. The Knight shift decreases below $T_c$ down to 20 K, followed by a still sharper drop below 20 K. Although the Knight shift decreases to almost zero, indicative of spin-singlet pairing, the temperature dependence of the Knight shift is quite different from the behavior seen in conventional spin-singlet superconductors. From the temperature dependences of both the Knight shift and $1/T_1$ at the F site in the superconducting state, Matano {\it et al.} suggested that there are two gaps opening below $T_c$, which possess nodes in the gap function\cite{MatanoEPL2008}. They suggested that a similar two-gap model is applicable to the analysis of the temperature dependence of $1/T_1$ below $T_c$ in LaFeAs(O$_{0.92}$F$_{0.08}$) ($T_c$=23 K)\cite{SKawasakiTwoGap}. 

\begin{figure}[tbp]
\begin{center}
\includegraphics[width=6.5cm]{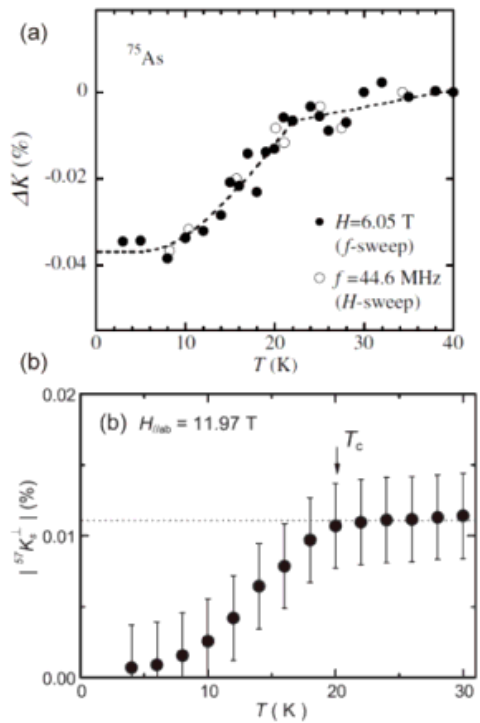}
\end{center}
\caption{(a) Temperature dependence of Knight shift at the As-site. The Knight-shift value is normalized at 40 K\cite{KawabataJPSJ08}. Figure reprinted from A. Kawabata {\it et al.}: J. Phys. Soc. Jpn {\bf 77} (2008) 103704. Copyright 2009 by the Japanese Physical Society (b) Temperature dependence of the spin-part of the Knight shift $|^{57}K_s^{\perp}|$ at the Fe site. The spin part of the Knight shift is evaluated from temperature dependences of the Knight shift at various sites\cite{TerasakiFeNMR}. Figure reprinted from N. Terasaki {\it et al.}: J. Phys. Soc. Jpn {\bf 78} (2009) 013701. Copyright 2009 by the Japanese Physical Society.}
\label{KSKawabata}
\end{figure}

Kawabata {\it et al.} and Terasaki {\it et al.} reported temperature dependences of Knight shift that are different from that reported by Matano {\it et al}\cite{KawabataJPSJ08,TerasakiFeNMR}. Figure \ref{KSKawabata}(a) shows the temperature dependence of the $^{75}$As Knight shift below $T_c$ measured by using a powder sample of LaFeAs(O$_{0.89}$F$_{0.11}$) ($T_c \sim$ 26 K)\cite{KawabataJPSJ08}. In contrast to that shown in Fig. \ref{KSMatano}, Knight shift decreases below $T_c$, but no additional decrease in the Knight shift suggesting a two-gap opening was observed. A similar temperature dependence of the Knight shift in the superconducting state was observed at the Fe site in LaFeAsO$_{0.7}$ with $T_c = 28$ K, as shown in Fig. \ref{KSKawabata}(b) \cite{TerasakiFeNMR}. Terasaki {\it et al.} estimated the spin part of the Knight shift by comparing the Knight shifts at various different sites, and showed that the spin-part of the Knight shift decreases to almost zero at $T = 0$, indicative of the spin-singlet pairing state.
The spin-singlet pairing state was also confirmed from the measurements of Knight-shift in different directions in Ba(Fe$_{1-x}$Co$_x$)$_2$As$_2$\cite{}.

\subsubsection{Isotope effect}
\begin{figure}[tb]
\begin{center}
\includegraphics[width=7cm]{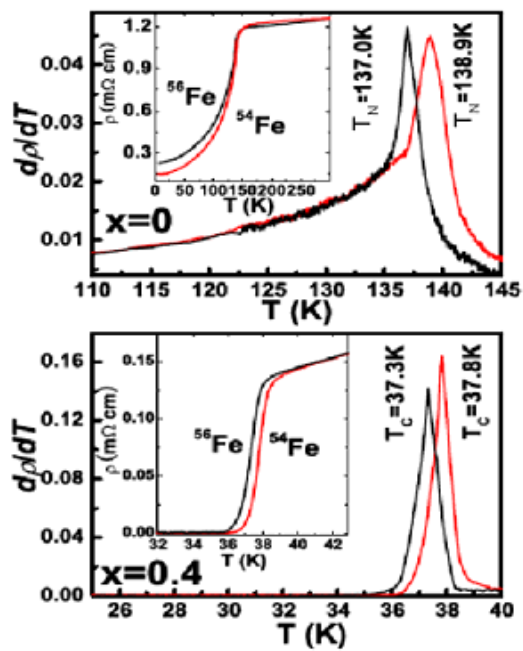}
\end{center}
\caption{(Color online) Temperature dependences of resistivity ($\rho$) and its derivative ($d\rho/dT$) for Ba$_{1-x}$K$_x$Fe$_2$As$_2$ ($x=0$ and 0.4) with $^{56}$Fe and $^{54}$Fe. A clear up-shift of $T_{\rm N}$ and $T_c$ is observed by replacing $^{56}$Fe with the isotope $^{54}$Fe. $\Delta T_{\rm N} = 1.9$K and $\Delta T_c = 0.5$K\cite{Isotope}.}
\label{Isotope}
\end{figure}
It is widely accepted that isotope effects on $T_c$ provide a strong evidence of the superconducting mechanism that electron-phonon interaction is responsible for superconductivity in the framework of the BCS theory. Liu {\it et al.} reported the effect of isotropic substitution on $T_c$ and an antiferromagnetic transition temperature $T_{N}$ in SmFeAsO$_{1-x}$F$_x$ by replacing $^{16}$O with isotope $^{18}$O and Ba$_{1-x}$K$_x$Fe$_2$As$_2$ by replacing $^{56}$Fe with isotope $^{54}$Fe\cite{Isotope}. Although $T_{N}$ and $T_c$ were found to change by approximately 0.61 and 0.75\% for oxygen isotope exchange in SmFeAsO$_{1-x}$F$_x$ with $x = 0$ and 0.15, these changes are 1.39 and 1.34\% for Fe isotope exchange in Ba$_{1-x}$K$_x$Fe$_2$As$_2$ with $x=0$ and 0.4, respectively, as shown in Fig.~\ref{Isotope}. The effect of the oxygen isotope is much smaller than that of the iron isotope, which suggests that the FeAs layer is a conducting layer responsible for superconductivity. An iron isotope component $\alpha$ was estimated to be about 0.4, being close to 0.5 for the full isotope effect in the framework of the BCS theory. These results seem to suggest that the electron-phonon interaction plays an important role in superconductivity of the pnictide superconductors. Since these results are crucial for understanding the mechanism of iron pnictide superconductivity, further experiments showing the variation in $T_c$ in several experimental runs are highly desired.

As for the local lattice structure, an extended X-ray absorption fine structure (EXAFS) measurement\cite{ZhangPRB08} revealed that the mean-square relative displacements of the Fe-As bond in LaFeAs(O$_{0.93}$F$_{0.07}$) exhibit a sharp drop at the $T_c$ onset, implying a significant electron-lattice interaction.

\subsubsection{Relationship between superconductivity and crystal structure}
\begin{figure}[tbp]
\begin{center}
\includegraphics[width=8cm]{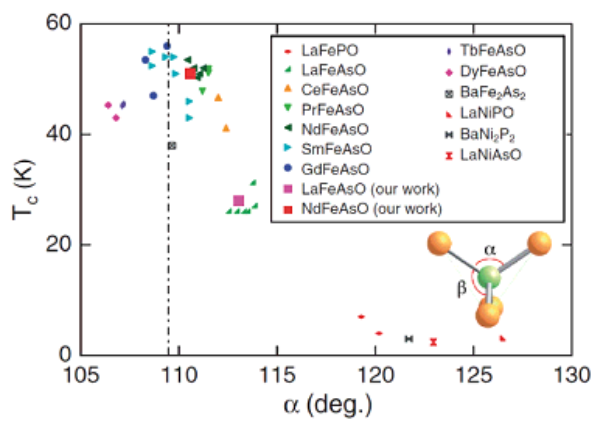}
\end{center}
\caption{(Color online) $T_c$ vs As-Fe-As bond angle $\alpha$ for various pnictide superconductors. The formulas of parent compositions of superconductors are depicted in the inset. The crystal structure parameters of samples showing almost maximum $T_c$ in each system are selected. The vertical dashed line indicates the bond angle of a regular tetrahedron ($\alpha=109.4^{\circ}$)\cite{LeeJPSJ2008}. Figure reprinted from C.-H. Lee {\it et al.}: J. Phys. Soc. Jpn {\bf 77} (2008) 083704. Copyright 2009 by the Japanese Physical Society}
\label{LeeStructure}
\end{figure}
Lee {\it et al.} have investigated the crystal structures of $R$FeAsO$_{1-\delta}$ ($R =$ La and Nd) by powder neutron diffraction analysis\cite{LeeJPSJ2008}. Atomic parameters of $R$FeAsO$_{1-\delta}$ were determined by Rietveld refinements of the neutron powder diffraction data, and the selected crystal structural parameters were obtained from the estimation of the bond lengths. The La and Nd series differ in a doping dependence of the thickness of the FeAs layer, which is determined by the distance between the As sites and the plane through the Fe sites. This As-Fe$_{\rm plane}$ distance increases more rapidly in the Nd-based compounds than in the La-based compounds. The As-Fe-As bond angle $\alpha$ of the Nd-based compounds, which is shown in the inset of Fig.~\ref{LeeStructure}, varies more rapidly with increasing oxygen deficiency. The FeAs$_4$ tetrahedron approaches a regular tetrahedron in which the angle $\alpha$ is 109.47$^{\circ}$. Figure \ref{LeeStructure} shows the relationship between the As-Fe-As bond angle $\alpha$ and $T_c$ in various pnictide superconductors. The parameters of the samples showing an almost maximum $T_c$ in each system are selected to eliminate the effect of carrier doping. It was revealed that $T_c$ becomes maximum when the FeAs$_4$ lattice forms a regular tetrahedron. A similar conclusion was also derived by Zhao {\it et al.}. They found a good correlation between $T_c$ and Fe-As/P-Fe angle in the ``1111'' compounds\cite{CeFeAsOPhaseDiagram}, suggesting that the structural perfection of the Fe-As tetrahedron is important for high-$T_c$ superconductivity.
These findings suggest that local symmetry around the Fe and As is the crucial parameter controlling $T_c$.

\section{Theoretical models}
Up to now, many theoretical studies of the iron-pnictide superconductors have been reported. As for the superconducting pairing state, most of them have proposed the $s_{\pm}$-pairing state. Here, we introduce two early works suggesting the $s_{\pm}$-pairing state, which were reported just after the discovery of the superconductors 

Mazin {\it et al.} argued that superconductivity realized in the iron-pnictide compounds is unconventional and mediated by antiferromagnetic spin fluctuations. Its pairing state is an extended $s$-wave pairing with a sign reversal of the order parameter between different Fermi surface sheets\cite{MazinPRL2008}. They claimed that doped LaFeAsO represents the first example of multigap superconductivity with a discontinuous sign change in order parameter between the bands, which is principally different from the multi-band $s$-wave superconductivity discovered in MgB$_2$. Their scenario is based on the calculated Fermi surfaces for undoped LaFeAsO, which is shown in Fig. 6, and the superconductivity is induced by the nesting-related antiferromagnetic spin fluctuations near the wave vectors connecting the electron and hole pockets. 

\begin{figure}[tbp]
\begin{center}
\includegraphics[width=7cm]{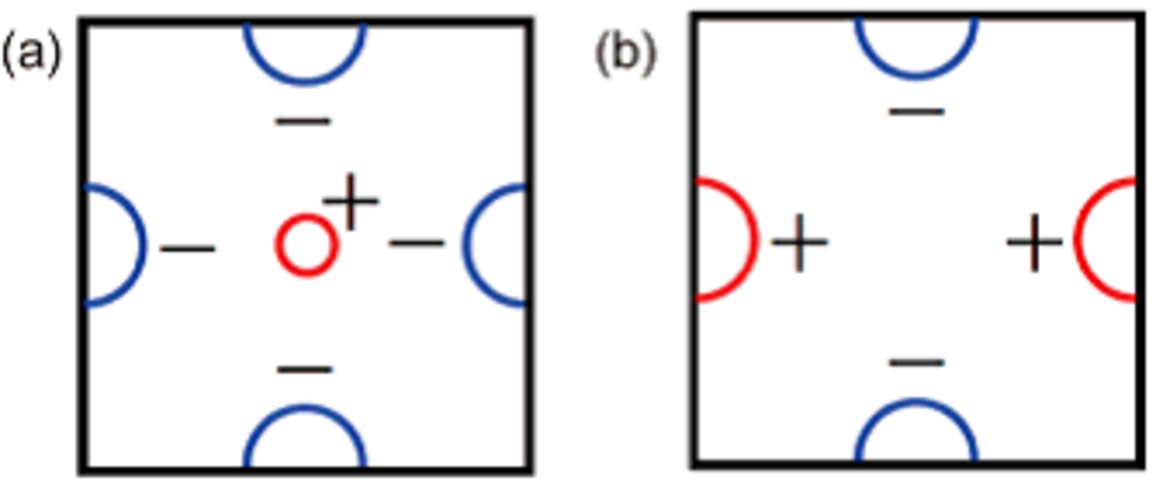}
\end{center}
\caption{(Color online) Schematical figures of (a) a fully gapped extended $s$-wave (called $s_{\pm}$-wave) and (b) $d_{x^2-y^2}$-wave gaps\cite{KurokiPRL2008}. Figure reprinted from K. Kuroki {\it et al.}: Phys. Rev. Lett {\bf 101} (2008) 087004. Copyright 2008 by the American Physical Society.}
\label{Kuroki}
\end{figure}

Kuroki {\it et al.} constructed a minimal model, where all the necessary five $d$-bands are included, and calculated spin and charge susceptibilities within random phase approximation\cite{KurokiPRL2008}. Furthermore, they investigated superconducting properties using the linearized Eliashberg equation, and concluded that the multiple spin-fluctuation modes arising from the nesting across the disconnected Fermi surfaces realize an extended $s$-wave pairing, in which the gap changes sign between the hole and electron Fermi surfaces across the nesting vector. This unconventional $s$-wave pairing is the same as the $s_{\pm}$ state proposed by Mazin {\it et al.}, which is schematically shown in Fig. \ref{Kuroki}(a). Kuroki {\it et al.} also suggest that a $d_{x^2-y^2}$-wave pairing, in which the gap changes sign between the electron Fermi surfaces as shown in Fig. \ref{Kuroki}(b), can also be another candidate, if the hole Fermi surfaces around $\Gamma$ are absent (or less effective). To identify the mechanism of iron-pnictide superconductivity, the determination of the presence or absence of nodes in the superconducting gap is quite important.

In the above pairing mechanisms, the antiferromagnetic fluctuations with the stripe-type correlation, which originate from the nesting between the hole- and electron Fermi surfaces, play an important role in superconductivity in the FeAs compounds. However, as shown in $\S 3.1.5$, the highest $T_c$ was observed in LaFeAs(O$_{1-x}$F$_x$) when the low-energy ($\sim$ mK order) antiferromagnetic fluctuations probed with the NMR measurements are suppressed over the entire $q$-space. The NMR results indicate the absence of magnetic fluctuations in the low-energy region in the high-$T_c$ samples and is considered to be incompatible with the above scenarios. However, when the characteristic magnetic fluctuations exceed the energy window of NMR measurements, the NMR measurements might be unable to detect the magnetic fluctuations with such high energies. Therefore, inelastic neutron-scattering measurements, which can investigate the energy and $q$ dependences of magnetic fluctuations, are important for fully understanding the relationship between magnetic fluctuations and superconductivity.

Quite recently, Ikeda has investigated the interplay between magnetic fluctuations and superconductivity in the effective five-band Hubbard model by the fluctuation-exchange approximation\cite{Ikeda}. It was shown by solving the superconducting Eliashberg equation that the most probable candidate for pairing symmetry is the sign-changed $s$-wave spin-singlet state. In addition, he showed that the eigenvalue is not so sensitive to carrier doping, and seems to be irrelevant with antiferromagnetic spin fluctuations if Hund's coupling $J$ is reasonably small. In this case, the correlation between $T_c$ and the antiferromagnetic fluctuations are weak and then $T_c$ is insensitive to carrier doping, in agreement with the experimental results. In his scenario, in addition to the magnetic fluctuations, the correlation between spins and spin quadrupoles cooperatively contributes to the pairing interaction.

\section{Summary}
On the basis of the above experimental results. we summarize the physical properties of iron-pnictide superconductors.

\begin{enumerate}
\item 
The parent compounds of iron-pnictide superconductors show metallic behavior in electric resistivity, differently from the insulating parent compounds of  cuprate superconductors. In addition, the tetragonal- orthorhombic transition and antiferromagnetic transition were observed at $T_S$ and $T_N$. In $R$FeAsO ($R$ : rare-earth elements), $T_S$ is slightly higher than $T_N$, and the magnetic transition seems to be induced by the structural transition. In the $A$Fe$_2$As$_2$ ($A$: Ba, Ca, and Eu), two transitions occur simultaneously with a first-order character. The antiferromagnetic $q$ vectors are $(\pi,0), (0,\pi)$ in the unfolded Brillouin zone, which corresponds to a stripe-type magnetic structure. The relationship between the structural transition and spin-density-wave type antiferromagntic transition is a key issue for understanding the mechanism of the magnetic order in the parent compounds. 

\item 
The magnetic ordering in the undoped ``1111'' compounds is suppressed by F-doping or oxygen deficiency, both of which correspond to electron doping. Superconductivity emerges when 3\% F is doped to the oxygen site in LaFeAs(O$_{1-x}$F$_x$), and $T_c$ is nearly unchanged up to 14 \% with a maximum $T_c \sim 26$ K at approximately F11\%. The ``122''-structure $A$Fe$_2$As$_2$ becomes superconducting with hole doping by the partial substitution of the $A$ site with monovalent $B^+$ (($A_{1-x}B_x$)Fe$_2$As$_2$) ($A$ = Ba, Sr, Ca, $B$ = K, Cs, Na), and by electron doping via partial substitution of cobalt for iron ($A$(Fe$_{1-x}$Co$_x$)$_2$As$_2$). Superconductivity was also observed in electron-doped $A$(Fe$_{1-x}$Co$_x$)AsF ($A$ = Ca and Sr) with the ``1111''structure. 

\item
The multiband properties of Fermi surfaces suggested by the band calculations have been confirmed by the ARPES measurements of single crystals of NdFeAs(O$_{1-x}$F$_x$) and (Ba$_{1-x}$K$_x$)Fe$_2$As$_2$. The characteristic wave vectors in the antiferromagnetic state seem to originate from the nesting between hole and electron Fermi pockets.

\item
The superconducting anisotropy of $\Gamma \equiv H_{c2}^{ab}/H_{c2}^{c}$ is rather small (4.3 and 2 in single-crystal NdFeAs(O$_{1-x}$F$_x$) and (Ba$_{1-x}$K$_x$)Fe$_2$As$_2$, respectively). This is quite different from the two-dimensional superconductivity observed in cuprate and organic superconductors.     

\item
The ARPES measurements of (Ba$_{1-x}$K$_x$)Fe$_2$As$_2$ reveal the presence of two superconducting gaps with different values: a larger gap ($\Delta\sim$12meV) on the small hole-like and electron-like Fermi surface sheets, and a smaller gap ($\Delta\sim$6meV) on the larger hole-like Fermi surface sheet. These gaps are nearly isotropic without nodes, and close simultaneously at the bulk $T_c$. The isotropic gap was also suggested from the measurements of penetration depth with microwave and $\mu$SR measurements.

\item
From the Knight-shift measurements in the superconducting state, the symmetry of the superconducting pairing is concluded to be in the spin-singlet state. $1/T_1$ in the superconducting state shows the absence of a coherence peak just below $T_c$ and the $T^3$ temperature dependence down to $\sim 0.2 T_c$. The $T^3$ dependence in the La-based ``1111'' compounds suggests highly anisotropic superconducting gaps with gap-zero regions along lines, which contradicts with the isotropic-gap state suggested from other experiments. The absence of the coherence peak in $1/T_1$ just below $T_c$ can be interpreted in terms of the $s_{\pm}$ wave model.

\item
From the $1/T_1T$ measurements of the undoped LaFeAsO and BaFe$_2$As$_2$, the stripe-type magnetic fluctuations revealed by the neutron-scattering experiments are shown. From the F concentration and temperature dependences of $1/T_1T$, the low-energy magnetic fluctuations are strongly suppressed by electron doping, and the pseudogap behavior is observed in $1/T_1T$ for LaFeAs(O$_{1-x}$F$_x$) and LaFeAsO$_{1-\delta}$ with the maximum $T_c$. $1/T_1T$ in LaFeAs(O$_{1-x}$F$_x$) significantly depends on F-concentration, whereas $T_c$ is insensitive to $x$ within a range from $x =$ 0.04 to 0.14. This suggests that the low-energy spin fluctuations detected at $1/T_1T$ do not play an important role in superconductivity of the iron-pnictide compounds. 

\item
From the $1/T_1T$ measurements of LaFeAs(O$_{1-x}$F$_x$), the pseudogap behavior becomes pronounced with F (electron)-doping. The doping dependence of the psuedogap behavior is opposite to that observed in the cuprate superconductors, in which the pseudogap behavior is significant in the underdoped region. It is considered that the pseudogap behavior is not ascribed to the antiferromagnetic correlations discussed in the cuprates, but to the character of the band structure near $E_{\rm F}$ in iron-pnictide superconductors.

\item
The neutron diffraction measurements revealed that $T_c$ becomes maximum when FeAs$_4$ forms a regular tetrahedron or when the Fe-As/P-Fe angle approaches 109.4$^{\circ}$. In addition, it was reported that the iron isotope component $\alpha$ was estimated to be approximately 0.4 (although more data showing the variation in $T_c$ in several experimental runs are highly desired), being close to 0.5 for the full isotope effect in the framework of the BCS theory, and that the oxygen-isotope effect is much smaller than the iron-isotope effect, indicative of the FeAs layer being responsible for superconductivity. These data suggest that the local geometry around Fe and As is related to superconductivity, particularly the local electron-phonon interaction originating from iron atoms.

\end{enumerate}

\section*{Acknowledgment}
One of the authors (Y. N.) is financially supported as a JSPS Research Fellow. 
The authors were supported by a Grant-in-Aid for ``Transformative Research-project on Iron Pnictides (TRIP)'' from Japan Science and Technology Agency (JST), a Grant-in-Aid for Scientific Research on Innovative Areas "Heavy Electrons" (No. 20102006) from The Ministry of Education, Culture, Sports, Science, and Technology (MEXT) of Japan, a Grant-in-Aid for the Global COE Program ``The Next Generation of Physics, Spun from Universality and Emergence'' from MEXT of Japan, and the Grants-in-Aid for Scientific Research from Japan Society for Promotion of Science (JSPS). 


\end{document}